\documentclass[a4paper, 11pt, oneside]{article}
\usepackage{aapreprint}
\makeatletter
\def\@fpheader{\today}
\makeatother

\usepackage{graphicx,wrapfig,float,slashed,cancel}
\usepackage{amsmath,amssymb,epsfig,graphicx,xcolor}
\usepackage{epstopdf}
\usepackage{soul}
\usepackage{ragged2e}
\usepackage{lipsum}
\usepackage{tabu}
\allowdisplaybreaks
\usepackage{pifont}
\usepackage[normalem]{ulem}
\usepackage[font={small,it},skip=0pt]{caption}

\newcommand{\nc}{\newcommand}
\nc{\non}{\nonumber}
\nc{\hc}{\hbox {h.c.}}
\nc{\noi}{\noindent}
\nc{\barx}{\bar{x}}
\nc{\pbarn}{\;\hbox {pb}}
\nc{\fbarn}{\;\hbox {fb}}
\nc{\hsp}{\hspace{0.5cm}}
\nc{\lsp}{\hspace{1cm}}
\nc{\Lsp}{\hspace{2cm}}
\nc{\LLsp}{\lsp\lsp}
\nc{\lra}{\longrightarrow}
\nc{\p}{\prime}
\nc{\sgn}{\text{sgn}}
\nc{\ph}{\varphi}
\nc{\op}{{\cal O}}
\nc{\eq}{\text{Eq.~}}
\nc{\cL}{\mathcal{L}}
\nc{\mo}{\texttt{micrOMEGAs }}
\nc{\vmol}{v_{\text{M\o l}}}
\nc{\eff}{\text{eff}}
\nc{\sm}{\text{\rm SM}}
\nc{\mchi}{m_\chi}
\nc{\mtchi}{m_{\tilde\chi}}
\nc{\mphi}{m_\phi}
\nc{\mtphi}{m_{\tilde\phi}}
\def\vx{v_{\textsc{x}}}
\def\gx{g_{\textsc{x}}}
\def\yx{y_{\textsc{x}}}

\nc{\beq}{\begin{equation}}  \nc{\eeq}{\end{equation}}
\nc{\bea}{\begin{eqnarray}}  \nc{\eea}{\end{eqnarray}}
\nc{\baa}{\begin{array}}     \nc{\eaa}{\end{array}}
\nc{\bit}{\begin{itemize}}   \nc{\eit}{\end{itemize}}
\nc{\ben}{\begin{enumerate}} \nc{\een}{\end{enumerate}}
\nc{\bce}{\begin{center}}    \nc{\ece}{\end{center}}
\nc{\bpm}{\begin{pmatrix}}   \nc{\epm}{\end{pmatrix}}
\nc{\bvt}{\begin{verbatim}}  \nc{\evt}{\end{verbatim}}
\def\lsim{\mathrel{\raise.3ex\hbox{$<$\kern-.75em\lower1ex\hbox{$\sim$}}}}
\def\gsim{\mathrel{\raise.3ex\hbox{$>$\kern-.75em\lower1ex\hbox{$\sim$}}}}
\def\udots{\mathinner{\mkern1mu\raise1pt\vbox{\kern7pt\hbox{.}}\mkern2mu\raise4pt\hbox{.}\mkern2mu\raise7pt\hbox{.}\mkern1mu}}
\def\kev{\;\hbox{keV}}
\def\mev{\;\hbox{MeV}}
\def\gev{\;\hbox{GeV}}

\title{\huge Multi-Component Dark Matter:\\ the vector and fermion case}
\author[1,2]{Aqeel~Ahmed,}
\author[1]{Mateusz~Duch,}
\author[1]{Bohdan~Grzadkowski,}
\author[1]{Michal~Iglicki}
\affiliation[1]{Faculty of Physics,
University of Warsaw,
Pasteura 5, 02-093 Warsaw, Poland}
\affiliation[2]{PRISMA Cluster of Excellence \& Mainz Institute for Theoretical Physics,\\ Johannes Gutenberg University, 55099 Mainz, Germany}
\emailAdd{aqeel.ahmed@fuw.edu.pl}
\emailAdd{mateusz.duch@fuw.edu.pl}
\emailAdd{bohdan.grzadkowski@fuw.edu.pl}
\emailAdd{michal.iglicki@fuw.edu.pl}

\abstract{
Multi-component dark matter scenarios constitute natural extensions of standard single-component setups and offer attractive new dynamics that could be adopted to solve various puzzles of dark matter. In this work we present and illustrate properties of a minimal UV-complete vector-fermion dark matter model where two or three dark sector particles are stable. The model we consider is an extension of the Standard Model (SM) by spontaneously broken extra $U(1)_X$ gauge symmetry and a Dirac fermion. All terms in the Lagrangian which are consistent with the assumed symmetry are present, so the model is renormalizable and consistent. To generate mass for the dark-vector $X_\mu$ the Higgs mechanism with a complex singlet $S$ is employed in the dark sector. Dark matter candidates are the massive vector boson $X_\mu$ and two Majorana fermions $\psi_\pm$. All the dark sector fields are singlets under the SM gauge group. The set of three coupled Boltzmann equations has been solved numerically and discussed. We have performed scans over the parameter space of the model implementing the total relic abundance and direct detection constraints. The dynamics of the vector-fermion dark matter model is very rich and various interesting phenomena appear, in particular, when the standard annihilations of a given dark matter are suppressed then the semi-annihilations, conversions and decays within the dark sector are crucial for the evolution of relic abundance and its present value. Possibility of enhanced self-interaction has been also discussed. 
}


\preprint{MITP/17--065}
\arxivnumber{1710.01853}
\begin{document}

\maketitle
\flushbottom

\section{Introduction}
\label{introduction}

The experimental data from the  WMAP \cite{Hinshaw:2012aka} and more recently
the Planck \cite{Ade:2015xua} collaborations provided an independent and indisputable confirmation for 
the presence of dark matter (DM) in the Universe. 
Nevertheless, in spite of a huge theoretical and experimental effort, its nature is still unknown. 
Till now only gravitational interactions of DM have been detected in a series of independent 
observations like the flatness of rotation curves of spiral galaxies \cite{Sofue:2000jx}, 
gravitational lensing \cite{Bartelmann:1999yn}, and collision of galaxy clusters with its 
pronounced illustration known as the Bullet cluster \cite{Clowe:2003tk}.
All attempts to detect DM non-gravitational interactions with ordinary matter have failed so far implying more and more stringent limits on
DM-nucleon cross-section, see e.g. LUX \cite{Akerib:2016vxi} and XENON-1T \cite{Aprile:2012nq} results.
The most popular models of DM are based on the assumption that it is composed of weakly interacting massive particles (WIMPs). Unfortunately, it turns out that the WIMP scenarios suffer from various difficulties when confronted with observations on small cosmological scales. 
For instance, the ``too-big-to-fail'' \cite{BoylanKolchin:2011de,Garrison-Kimmel:2014vqa} and the ``cusp-core'' \cite{Moore:1994yx,Flores:1994gz,Oh:2010mc,Walker:2011zu} problems are widely discussed in  the literature.  In particular, the DM densities inferred in the central regions of DM dominated galaxies are usually smaller than expected from WIMP simulations \cite{Rocha:2012jg,Weinberg:2013aya}. 
It turns out that an appealing solution to those problems is to assume that dark matter self-interacts strongly \cite{Spergel:1999mh}. The assumption of self-interacting dark matter (SIDM) implies that central (largest) DM density could be reduced. Usually, self-interacting DM scenarios require the presence of light DM and also light DM mediators.

Dark matter could also be searched for through indirect detection experiments, which assume that in regions of large DM density, its pairs are likely to annihilate. Then secondary particles released in this process, e.g.  gamma rays, neutrinos, electrons, positrons, protons and anti-protons, and could be observed on Earth, which could reveal some properties of DM. Independent analysis of the Fermi-LAT data \cite{Atwood:2009ez} by various groups have shown an excess of gamma ray in the energy range $1\!-\!3\gev$
that can be interpreted as a result of DM annihilation in the Galactic Center. Besides the $1\!-\!3\gev$ excess gamma rays,
there exists an observation of unidentified $3.55\kev$ X-ray line found by \cite{Boyarsky:2014jta} and \cite{Bulbul:2014sua}.
As shown by several groups, this unknown X-ray line can also be explained by DM annihilation. To explain the indirect signals relatively large DM mass is needed, e.g. $\sim 50\gev$ in \cite{Calore:2014nla}.

Since very different DM masses are needed to solve the small-scale problems (through self-interaction) and to interpret the potential indirect signals, therefore in order to accommodate both observations, a multi-component DM seems to be a natural option. 
Various multi-component DM models have been proposed and studied in the literature, for instance, multi-scalar DM~\cite{Grzadkowski:2009mj,Grzadkowski:2009bp,Drozd:2011aa,Grzadkowski:2011jks,Belanger:2012vp,Ivanov:2012hc,Modak:2013jya,Aoki:2013gzs,Biswas:2013nn,Drozd:2013aea,Biswas:2015sva,Bhattacharya:2016ysw,Bhattacharya:2017fid}, multi-fermion DM~\cite{Cao:2007fy,Huh:2008vj,Fukuoka:2010kx,Cirelli:2010nh,Heeck:2012bz,Belanger:2011ww,Kajiyama:2013rla,Gu:2013iy,Bell:2013wua}, multi-vector DM~\cite{Gross:2015cwa,Karam:2016rsz}, scalar-fermion DM~\cite{DEramo:2010keq,Daikoku:2011mq,Aoki:2012ub,Bhattacharya:2013hva,Bhattacharya:2013asa,Bae:2013hma,Aoki:2014lha,Esch:2014jpa,DuttaBanik:2016jzv,Khan:2017ygl}, scalar-vector DM~\cite{Bian:2013wna,Bian:2014cja,Arcadi:2016kmk}, vector-fermion DM~\cite{DiFranzo:2016uzc,Ma:2017ucp}, 
and various other generic multi-component DM~\cite{Zurek:2008qg,Feng:2008ya,Profumo:2009tb,Batell:2010bp,Feldman:2010wy,Dienes:2011ja,Dienes:2011sa,Chialva:2012rq,Geng:2013nda,Dienes:2014via,Geng:2014zea,Geng:2014dea,Buen-Abad:2015ova,Lai:2015edx,Geng:2015qth,Dienes:2016kgc,Dror:2016rxc,Dienes:2016vei} scenarios. 
Models of multi- component DM were also considered and adopted in astrophysical simulations, e.g. by 
\cite{Malekjani:2007pv,Semenov:2013vya,Medvedev:2013vsa,Demianski:2015pna}. Needless to say, the dynamics of multi-component DM
is much richer than that of a simple WIMP, and therefore attractive to study by itself, even without any phenomenological direct application. In particular multi-component DM models allow to have, besides the standard annihilations and co-annihilations; conversion, semi-annihilation, and decay processes which make the dark sector (thermal) dynamics much more involved and interesting. Note that most of the models mentioned above, discuss the implications of one or two of these multi-component DM properties.

In this work, we propose a minimal UV complete vector-fermion DM model that predicts two or three stable dark states. Our model involves an extension of the SM by an Abelian dark gauge symmetry $U(1)_X$. The model is minimal in a sense that it contains only three new fields in the dark sector; a dark gauge boson $X_\mu$, a Dirac fermion $\chi$, and a complex scalar $S$, which serves as a Higgs field in the hidden sector. They are singlets under the SM gauge group but they are all charged under the dark $U(1)_X$ gauge symmetry and therefore they interact with each other. The complex scalar $S$ acquires a vacuum expectation value (vev) and gives mass to the dark gauge boson $X_\mu$ by the dark sector Higgs mechanism. It also contributes to the mass of the Dirac fermion $\chi$ through the Yukawa coupling. Moreover, the presence of the Dirac mass for the fermion introduces a mixing of its chiral components. After diagonalization of the mass matrix, the Dirac fermion splits into two Majorana fermions $\psi_\pm$ with mass eigenvalues $m_\pm$. As a result, after the dark sector symmetry breaking we have three potentially stable interacting particles: a dark vector and two Majorana fermions. Their stability is ensured by a residual $Z_2\!\times\!Z_2^\p$ discrete symmetry, 
which also dictates the possible form of dark sector interactions. The communication with the visible (SM) sector proceeds only via the Higgs portal $\kappa |H^2||S^2|$.

Our minimal vector-fermion dark matter model has many attractive features. First of all, the very fact that in the multi-component DM literature, the vector-fermion dark matter possibility has not been studied in detail\,\footnote{As referred above there are only two recent works~\cite{DiFranzo:2016uzc,Ma:2017ucp} which consider the possibility of an Abelian vector and a fermion as two-component dark matter scenario. These models share some properties with the model analyzed here, however their fermionic sectors are different.} speaks for itself and therefore the goal of this work is to provide an extensive analyzes of the minimal vector-fermion scenario. Some of the interesting features of the model are: (\!{\it i}) the presence of a second scalar helps to achieve the stability of the SM Higgs potential even at the tree level, see e.g.~\cite{Lebedev:2012zw,Duch:2015jta,Duch:2015cxa}, (\!{\it ii}), a possibility of enhancing vector component self interactions, see e.g.~\cite{Duch:2017khv}, (\!{\it iii}) a very small/large mass splitting among the dark sector states (vector and Majorana fermions) are possible without large tuning of the parameters, (\!{\it iv}) our model is a gauged version of the model considered by Weinberg~\cite{Weinberg:2013kea} for different purposes, and (\!{\it v}) more importantly the minimality of the model; since there is only one parameter, the dark gauge coupling coupling $\gx$, which controls the dynamics in the dark sector, including the conversion, semi-annihilation and decay processes. In this work, we are going to illustrate the relevance of conversions, semi-annihilations, and decays in the vector-fermion DM model.

The paper is organized as follows. In sec.~\ref{Vector-fermion two-component dark matter} the vector-fermion (2-3 component) model of DM is presented. Solutions of three coupled Boltzmann equations are discussed in sec.~\ref{Boltzmann equations for VFDM} focusing on conversion, semi-annihilation, and decay processes. There we show results of a detailed scan over the parameter space of the model satisfying the observed total relic density and direct detection constraints. In section~\ref{selfintDM} we focus on the region with large self-interaction cross-section.
Section~\ref{conclusions} contains summary and conclusions. Moreover, we supplement our work with an~\ref{Boltzmann equations for semi-annihilating dark matter}, collecting details of the derivation of Boltzmann equations, and an~\ref{DD_multi_DM}, describing the method adopted to obtain constraints for a multi-component DM model by direct detection experiments.

\section{Vector-fermion multi-component dark matter model}
\label{Vector-fermion two-component dark matter}
 In this section, we explore the possibility of having a multi-component dark matter model with a vector boson and a Dirac fermion (charged under a dark sector gauge symmetry) which may serve as dark matter candidates. As mentioned in the Introduction, we consider a minimal extension of the SM by an additional $U(1)_X$ gauge symmetry of the dark sector with a complex scalar field $S$ and a Dirac fermion $\chi$, both charged under the dark-sector gauge group. We employ the Higgs mechanism in the dark sector such that the vev of the complex scalar $S$ generates a mass for the $U(1)_X$ gauge field $X_\mu$. 
The dark-segment fields have the following quantum numbers under the gauge group $SU(3)_c\!\times \!SU(2)_L\!\times \!U(1)_Y\!\times\! U(1)_X$,
\begin{align}
S&=(\mathbf1,\mathbf1,0,1),\lsp \chi=(\mathbf1,\mathbf1,0,\tfrac12).    \label{schi_qnum}
\end{align}
We assume that none of the SM fields is charged under the dark gauge symmetry $U(1)_X$.

We can write down the Lagrangian for our simplest vector-fermion MCDM model as
\begin{align}
\cL&=\cL_{\rm SM}+\cL_{\rm DS}+\cL_{\rm portal},  \label{L_full}
\end{align}
where $\cL_{\rm SM}$ is the SM Lagrangian, $\cL_{\rm DS}$ is the dark-sector Lagrangian,
\begin{align}
\cL_{\rm DS}&=-\frac14X_{\mu\nu} X^{\mu\nu}+\big({\cal D}_\mu S\big)^\ast {\cal D}^\mu S+\mu_S^2|S|^2-\lambda_S |S|^4 \notag\\
&\quad~+\bar\chi \big(i\slashed {\cal D}-m_D\big)\chi-\frac1{\sqrt2}\big(\yx S^\ast \chi^\intercal {\cal C}\chi+\hc\big) ,  \label{L_DS}
\end{align}
and $\cL_{\rm portal}$ is the Lagrangian for the Higgs portal interactions between the SM and the dark sector,
\begin{align}
\cL_{\rm portal}&=-\kappa|S|^2|H|^2.  \label{L_int}
\end{align}
Note that the portal coupling is the only communication between the SM and the dark sector. 
Above, in \eq\eqref{L_DS} $X_{\mu\nu}\equiv\partial_\mu X_\nu-\partial_\nu X_\mu$ is the field strength tensor for the $X_\mu$ field and ${\cal D}_\mu$ is the covariant derivative defined as
\beq
{\cal D}_\mu=\partial_\mu+i\,\gx \,q_{\textsc x}\, X_\mu,     \label{co_der_x}
\eeq
where $\gx$ is the coupling constant corresponding to the gauge group $U(1)_X$, whereas $q_{\textsc x}$ are the $U(1)_X$ charges $1$ and $\tfrac12$ for $S$ and $\chi$ (as defined in \ref{schi_qnum}), respectively. Moreover, in \eq\eqref{L_DS} $m_D$ is the Dirac mass, $\yx$ is the dark Yukawa coupling and ${\cal C}\!\equiv \! -i \gamma_2\gamma_0$ is the charge-conjugation operator, where $\gamma_0,\gamma_2$ are the gamma matrices. It is important to note that the dark sector is invariant with respect to the charge conjugation symmetry ${\cal C}$ under which the dark fields transform as follows:
\begin{align}
X_\mu \xrightarrow{{\cal C}} -X_\mu,\hsp
S \xrightarrow{{\cal C}} S^\ast, \hsp
\chi \xrightarrow{{\cal C}} \chi^c\equiv-i\gamma_2 \chi^\ast.     \label{c_symmetry}
\end{align}
The symmetry forbids a kinetic mixing between the weak hypercharge $U(1)_Y$ and the dark $U(1)_X$, so that $X_\mu$ can not decay into SM particles. 

It is useful to collect the full scalar potential of our model,
\beq
\begin{aligned}
V(H,S)=&-\mu_H^2 |H|^2-\mu_S^2 |S|^2	 +\lambda_H |H|^4+\lambda_S |S|^4+\kappa |H|^2 |S|^2. \label{potential}
\end{aligned}
\eeq
Tree-level positivity/stability of the above scalar potential requires the following conditions to be satisfied:
\beq
\lambda_H>0,\lsp \lambda_S>0, \lsp \kappa>-2\sqrt{\lambda_H\lambda_S}.  \label{positivity}
\eeq
It is straightforward to find the minimization conditions for the scalar potential \eqref{potential} as\
\beq
\begin{aligned}
\big(2\lambda_H v^2-2\mu_H^2+\kappa \vx^2\big)v &=0,   \lsp  &\big(2\lambda_S \vx^2-2\mu_S^2+\kappa v^2\big)\vx &=0,    \label{min_cond}
\end{aligned}
\eeq
where $ \langle H^\intercal\rangle\equiv (0,v/\sqrt2)$ and $\langle S\rangle\equiv \vx/\sqrt2$ are the vevs of the respective fields\footnote{Note that because of the $U(1)_X$ symmetry, $\vx$ can be chosen to be real and positive. Therefore the charge conjugation \eqref{c_symmetry} remains unbroken.}. In order to have both vevs non-zero ($v$ provides masses to the SM gauge bosons and the dark-sector scalar vev $\vx$ generates mass for the dark vector) we require $\kappa^2>4\lambda_H\lambda_S$ and the values of vevs are:
\beq
v^2=\frac{4\lambda_S\mu_H^2-2\kappa\mu_S^2}{4\lambda_H\lambda_S-\kappa^2},  \lsp
\vx^2=\frac{4\lambda_H\mu_S^2-2\kappa\mu_H^2}{4\lambda_H\lambda_S-\kappa^2}.  \label{vevs}
\eeq

We expand the Higgs doublet and the singlet around their vevs as:
\beq
H=\frac1{\sqrt2}\bpm \sqrt2 \pi^+\\ v+h+i\pi^0\epm, \lsp    S=\frac1{\sqrt2}\big(\vx+\phi+i\sigma\big),   \label{HS_pert}
\eeq
where $\pi^{0,\pm}$ and $\sigma$ are the would-be Goldstone modes that disappear in the unitary gauge. Hence, only the scalars $h$ and $\phi$ correspond to the physical states. The mass squared matrix for the scalar fluctuations $(h,\phi)$ is given by,
\beq
{\cal M}^2=\bpm 2\lambda_H v^2& \kappa v \vx \\ \kappa v \vx& 2\lambda_S \vx^2\epm. \label{mass_matrix}
\eeq
The above mass squared matrix ${\cal M}^2$ can be diagonalized by the orthogonal rotational matrix ${\cal R}$, such that,
\beq
\begin{aligned}
{\cal M}^2_{\text{diag}}\equiv {\cal R}^{-1}{\cal M}^2{\cal R}&=\bpm m_{h_1}^2 &0\\0&m_{h_2}^2\epm, \hsp
\text{where}\hsp &{\cal R}&=\bpm\cos\!\alpha & -\sin\!\alpha\\ \sin\!\alpha &\cos\!\alpha \epm,    \label{mass_matric_diag}
\end{aligned}
\eeq
and $(h_{1},h_{2})$ are the two physical mass-eigenstate Higgs bosons with masses $(m_{h_{1}}^2,m_{h_2}^2)$, defined in terms of $(h,\phi)$ as
\beq
\bpm h_1 \\ h_2\epm ={\cal R}^{-1}\bpm h\\ \phi\epm.  \label{higgs_physical}
\eeq
The mixing angle $\alpha$ could be expressed in terms of mass matrix elements as follows:
\beq
\tan(2\alpha)=\frac{\kappa  v \vx}{ v^2\lambda_H- \vx^2\lambda_S}\,.        \label{tan2a}
\eeq
We will later adopt $\sin\!\alpha$ as an independent input parameter while scanning over model parameters.
The masses for the two eigenstates $h_1$ and $h_2$ are 
\begin{align}
m_{h_1}^2&=v^2 \lambda_H + 
 \vx^2 \lambda_S + (v^2\lambda_H- \vx^2\lambda_S)/\!\cos(2\alpha)\,, \label{mh1sq}\\ 
 m_{h_2}^2&=v^2 \lambda_H + 
 \vx^2 \lambda_S - (v^2\lambda_H- \vx^2\lambda_S)/\!\cos(2\alpha)\,.    \label{mh2sq}
\end{align}
In this analysis we will treat the two Higgs masses as independent parameters. It is always assumed that $h_1$ is the observed state with $m_{h_1}\!=\!125\gev$. The other mass eigenstate could be either lighter or heavier than $h_1$.
After the spontaneous symmetry breaking (SSB) the SM fermions acquire mass from the SM Higgs doublet, whereas the dark sector fermion $\chi$ receives mass from the dark-scalar $S$ through the Yukawa interaction term and from the Dirac mass term. After the SSB the dark fermionic sector Lagrangian can be rewritten as, 
\begin{align}
{\cal L}_{\textsc{df}}=&\frac{i}2\big(\bar\chi \gamma^\mu \partial_\mu\chi+\bar{\chi^c} \gamma^\mu \partial_\mu\chi^c\big)-\frac{m_D}2\big(\bar\chi\chi+\bar{\chi^c} \chi^c\big)\notag \\
&-\frac{\yx\vx}{2}\big(\bar{\chi^c }\chi+\bar\chi\chi^c \big) -\frac{\gx}4 \big(\bar \chi \gamma^\mu\chi-\bar {\chi^c} \gamma^\mu\chi^c\big) X_\mu -\frac{\yx}{2}\big(\bar{\chi^c} \chi+\bar\chi\chi^c \big)\phi .  \label{L_FS}
\end{align}
In the above Lagrangian the field $\chi$ and its charge-conjugate $\chi^c$ mix through the Yukawa interactions. One can easily diagonalize the above Lagrangian in terms of the following Majorana mass-eigenstates $\psi_\pm$ ($=\psi_\pm^c$)
\beq
\psi_{\!+}\equiv \frac1{\sqrt2}\big(\chi+\chi^c\big), \lsp \psi_{\!-}\equiv \frac1{i\sqrt2}\big(\chi-\chi^c\big),     \label{psi_pm}
\eeq
with mass eigenvalues,
\beq
m_\pm=m_D\pm y \vx.
\eeq
In the new basis we can rewrite the above dark fermionic Lagrangian as, 
\begin{align}
{\cal L}_{\textsc{df}}=&\frac{i}2\big(\bar\psi_{\!+} \gamma^\mu \partial_\mu\psi_{\!+} +\bar\psi_{\!-} \gamma^\mu \partial_\mu\psi_{\!-}\big)-\frac12m_+\bar\psi_{\!+}\psi_{\!+} -\frac12 m_-\bar\psi_{\!-} \psi_{\!-}  	\notag \\
&-\frac{i}4\gx \big(\bar \psi_{\!+} \gamma^\mu\psi_{\!-} - \bar \psi_{\!-} \gamma^\mu\psi_{\!+}\big) X_\mu -\frac{\yx}{2}\big(\bar\psi_{\!+}\psi_{\!+} -\bar\psi_{\!-} \psi_{\!-} \big)\phi .  \label{L_FS_psi}
\end{align}
The mass splitting between $\psi_+$ and $\psi_-$ is controlled by the Yukawa coupling and $\vx$: 
\beq
\Delta m\equiv m_+-m_- =2 \yx \vx.
\eeq
Hereafter, we will assume $\yx \!>\! 0$ and since $\vx$ could be chosen positive therefore we have $m_+\!>\!m_-$. 

Note that the above Lagrangian is invariant with respect to a discrete symmetry $Z_2\!\times \!Z_2^\p$, under which the SM fields are even whereas the dark sector fields transform non-trivially, as given in Tab.~\ref{tab:Z2Z2p}. 
It is easy to see that the above $Z_2$ is a direct consequence of the charge conjugation symmetry 
\eqref{c_symmetry}. It is worth to notice that since $\vx$ is real (without compromising any generality) therefore the charge conjugation remains unbroken so that $X_\mu$ cannot decay into SM particles, see also \cite{Ma:2017ucp}. 
On the other hand $Z_2^\p$ is implied by the $U(1)_X$ gauge symmetry. Note that the $Z_2^{\p\p}$ is responsible for the stability of $\psi_-$ since it is lighter than $\psi_+$. It is interesting to notice that our model has also a discrete $Z_4$ symmetry~\footnote{For generic discussion on $Z_N$ discrete symmetries as the residual of an Abelian gauge symmetry see~\cite{Batell:2010bp}.} under which the three dark matter components are charged. The $Z_4$ charges are: 
\beq
\Phi\to e^{i\pi n_{\Phi}} \Phi, \lsp {\rm where}\hsp \Phi=(X_\mu,\psi_\pm,\phi),
\eeq
with $n_\Phi=(1,\pm \tfrac12,0)$, whereas all the SM particles are neutral under this symmetry. However, for our analysis the $Z_2\!\times\!Z_2^\p$ (along with $Z_2^{\p\p}$) completely specifies all the relevant properties of $Z_4$, therefore hereafter we only consider $Z_2\!\times\!Z_2^\p$.  
\begin{table}[t]
\caption{Discrete symmetries of the vector-fermion DM model.} \label{tab:Z2Z2p}
\centering{\tabulinesep=1pt
\rowcolors{1}{gray!23}{white}  
\begin{tabu}to 9cm{|[1pt]X[2,c]|[0.75pt]X[1,c]|[1pt]X[1,c]|[0.75pt]X[1,c]|[0.75pt]X[1,c]|[1pt]}\hline
Symmetry &$X_\mu$ & $\psi_{\!+}$&$\psi_{\!-}$ & $h_i({\rm SM})$\\
\hline
$Z_2$ &$-$& $+$ &$-$ &$+$\\
$Z_2^\p$ &$-$ & $-$ &$+$ &$+$\\
$Z_2^{\p\p}$ &$+$ & $-$ &$-$ &$+$\\
\hline
\end{tabu}
}
\end{table}
\begin{figure}[t]
\includegraphics[width=\textwidth]{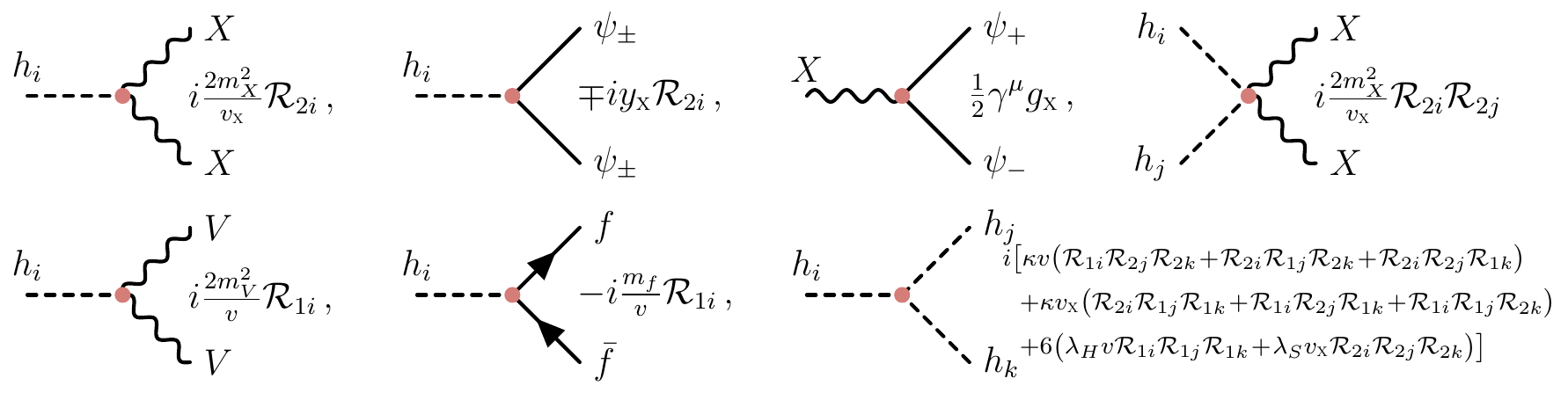}
\caption{The most relevant Feynman rules for vector-fermion dark matter sector ($X$ is the dark vector and $\psi_\pm$ are the dark fermions) and its mixing with the SM, where $V$ represents the SM gauge bosons ($Z,W$), $f$ denotes the SM fermions, and ${\cal R}$ is the rotation matrix, defined in Eq.~\eqref{mass_matric_diag}, where the two scalars $h_1,h_2$ mix with mixing angle $\alpha$.}
\label{feynman_rules}
\end{figure} 

As it can be seen from the above fermionic Lagrangian there is only one interaction term between all three components of the dark segment (the vector boson $X_\mu$ and the two Majorana fermions $\psi_\pm$):
\beq
-\frac{i}4\gx \big(\bar \psi_{\!+} \gamma^\mu\psi_{\!-} - \bar \psi_{\!-} \gamma^\mu\psi_{\!+}\big) X_\mu\,. \label{dark_vertex}
\eeq
In Fig.~\ref{feynman_rules}, we collect all the relevant vertices that desckineticribe interactions of the dark segment of the theory. 

As depicted in Fig.~\ref{vfdm_classification}, depending on the masses of dark particles, there are three cases~\footnote{Since we have assumed $m_-\!<\!m_+$, hence there are only two 2CDM cases.} in which two or all three particles could be stable and serve as dark matter: 
\begin{figure}[t]
\centering
\includegraphics[width=0.67\columnwidth]{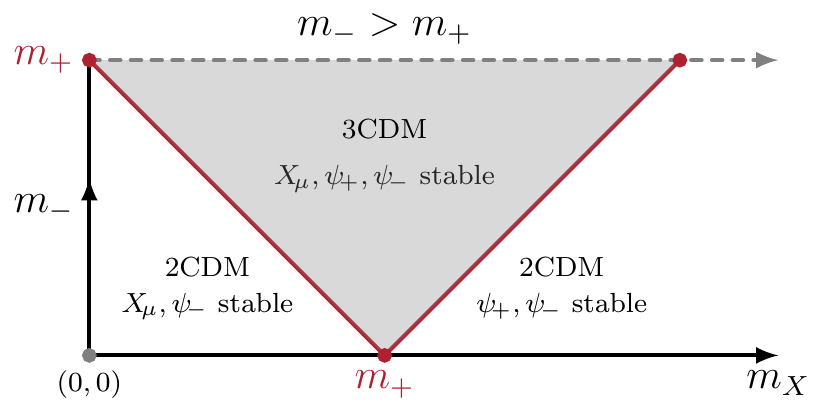}
\caption{Schematic diagram for the 2- and 3-component vector-fermion dark matter scenarios. We consider $m_+$ to have a fixed value and the gray region represent parameter space where the all three dark sector particles ($X_\mu,\psi_+,\psi_-$) are stable, whereas the white regions represent the 2-component scenarios where $\psi_+$ and $X_\mu$ are unstable, respectively.}
\label{vfdm_classification}
\end{figure}
\bit\itemsep0em
\item The first case is when $m_+ \!>\! m_-+m_X$, the Majorana fermion $\psi_{\!+}$ will decay into a stable vector $X_\mu$ and a stable Majorana fermion $\psi_{\!-}$. This is a 2CDM case, the white area (left) in Fig.~\ref{vfdm_classification}.
\item The second case is when $m_X\!>\! m_++m_-$, the  vector $X_\mu$ will decay into two stable Majorana fermions $\psi_\pm$. This is a 2CDM case, the white area (right) in Fig.~\ref{vfdm_classification}.
\item The third case is when $m_+ \!+\!m_-\!>\!m_X\!>\!m_+\!-\!m_-$, so that none of the three particle will decay and hence all are stable. This is a 3CDM case, shown as gray region in Fig.~\ref{vfdm_classification}. Note that the boundaries (right/left) of the gray region correspond to the case when $m_X=m_+\pm m_-$. 
\eit

\subsection{Input parameters}
\label{Input parameters}
Here we outline the strategy adopted to investigate the vector-fermion dark matter (VFDM) model. First of all, we would like to count the free parameters in the model and the number of constraints. There are five parameters in the scalar potential \eqref{potential}, namely, mass parameters $\mu_H,\,\mu_S$, quartic couplings $\lambda_H,\lambda_S$ and the portal coupling $\kappa$. Additionally, there are three parameters in the dark gauge and fermionic part, i.e. the dark gauge coupling $\gx$, the fermion Dirac mass $m_D$, and the Yukawa coupling $\yx$, see Eq.~\eqref{L_DS}. In total there are eight parameters of the model, however there are two constraints from the SM Higgs vev $v$ and the Higgs mass $m_{h_1}$, hence leaving six parameters free. 
We adopt the following set as an independent input parameters:  
\beq
m_{h_2},\,m_X,\, m_+,\, m_-,\, \sin\!\alpha,\, \gx.
\eeq
Note that the mixing angle $\sin\!\alpha$ is constrained by the SM-like Higgs couplings with the gauge bosons. We employed $|\sin\!\alpha|\!\leq\!0.33$, the current LHC $2\sigma$ bound \cite{Khachatryan:2016vau}.

Remaining parameters could be expressed in terms of the input set as follows (note the potential mass parameters $\mu_H,\,\mu_S$ are already traded for $v,\, \vx$, see Eq.~\ref{vevs}):
\begin{align}
\vx&=\frac{m_X}{\gx}\,,  \lsp &\kappa&=\frac{(m_{h_1}^2 -m_{h_2}^2) \sin (2 \alpha )}{2v\vx}\,, \\
\lambda_H&=\frac{m_{h_1}^2\cos^2\alpha+m_{h_2}^2\sin^2\alpha}{2 v^2}\,,   &\lambda_S&=\frac{m_{h_1}^2\sin^2\alpha+m_{h_2}^2\cos^2\alpha}{2 \vx^2}\,, \\
\yx&=\frac{(m_+-m_-)}{2 \vx}\,, \lsp &m_D&=\frac{(m_++m_-)}{2}\,.
\end{align}
Note that the Yukawa coupling $\yx\!=\!\Delta m/(2 \vx)\! =\! \Delta m/(2m_X)\!\times \!\gx$, therefore for fixed $m_X$ the Yukawa coupling $\yx$ is proportional to $\gx$. In other words, for fixed $m_X$, the vev $\vx$ must vary if $\gx$ is being changed, implying a variation of $\yx$. This remark is important hereafter, e.g. for fixed $\Delta m$ and $m_X$ one has to adjust $\gx$ in order to satisfy constraints from direct detection experiments even in the case when DM is dominated by $\psi_\pm$.

\section{Vector-fermion dark matter phenomenology}
\label{Boltzmann equations for VFDM}
\begin{figure}[t]
\includegraphics[width=\textwidth]{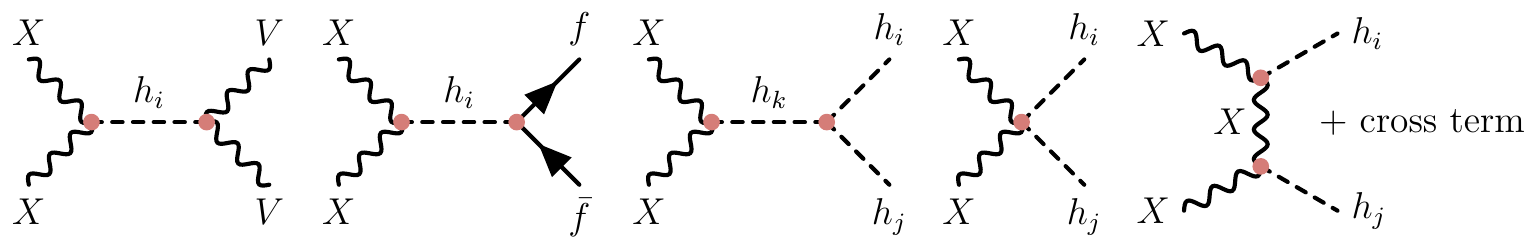}\newline
\includegraphics[width=\textwidth]{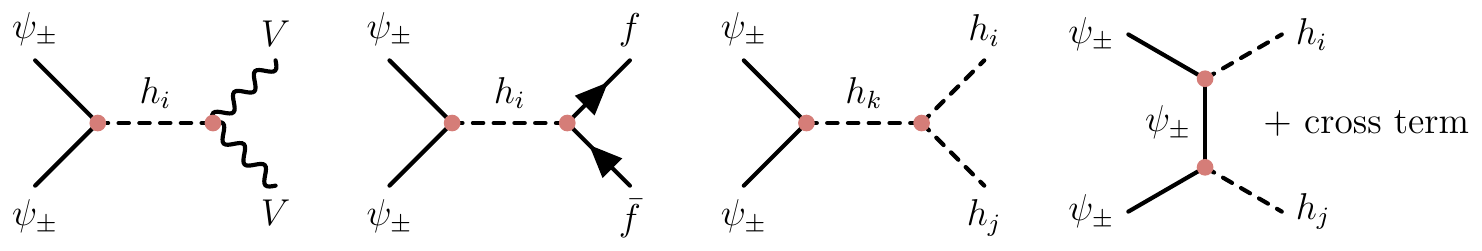}
\caption{The vector dark matter $X_\mu$ and Majorana fermion dark matter $\psi_\pm$ annihilation diagrams. Above $V$ and $(\bar f)f$ denote the SM vector bosons ($W^\pm$ and $Z$) and the SM (anti)fermions (quarks and leptons).}\label{fig:annihilation}
\end{figure}

\begin{figure}[t]
\includegraphics[width=\textwidth]{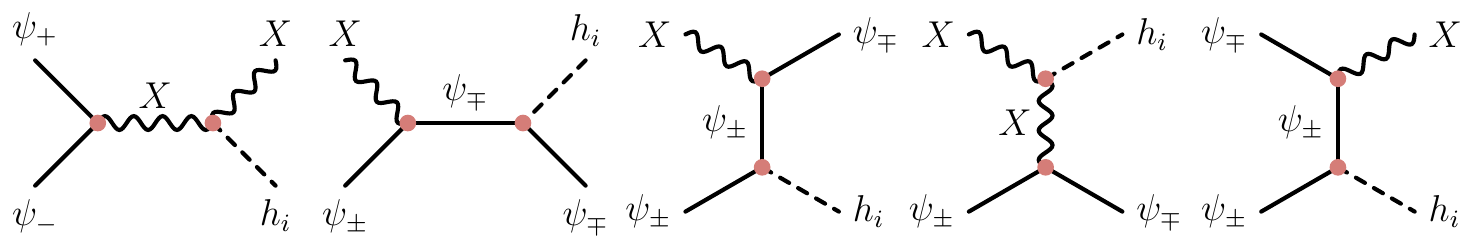}
\caption{Semi-annihilation diagrams for the dark particles $X, \psi_\pm$. In these processes two of the dark sector particles annihilate to a dark sector particle and a SM particle in the thermal bath.}\label{fig:semi_annihilation}
\end{figure}

\begin{figure}[t]
\includegraphics[width=\textwidth]{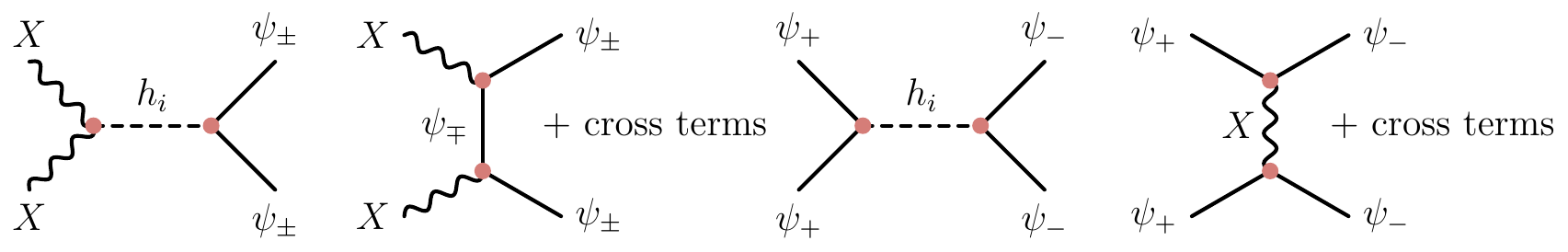}
\caption{Dark matter conversion processes involving $X$ and  $\psi_\pm$. These processes are important to keep thermal equilibrium within the dark sector.}\label{fig:conversion}
\end{figure}

In this section we present coupled Boltzmann equations for the evolution of number density of dark matter particles ($X_\mu,\psi_+,\psi_-$) in VFDM model. Figures~\ref{fig:annihilation}, \ref{fig:semi_annihilation}~and~\ref{fig:conversion} contain Feynman diagrams relevant for collision terms for the annihilation, semi-annihilation and conversion processes, respectively. It is assumed that dark matter components scatter against SM particles frequently enough so that their temperatures are the same as that of the thermal bath. 

The Boltzmann equations for the DM components ($X_\mu, \psi_+,\psi_-$), can written as:
\begin{align}
\frac{dn_{X}}{dt}=& -3H n_{X}-\langle\sigma^{XX\phi\phi^\p}_{\vmol}\rangle\Big(n_{X}^2-\bar n_{X}^2\Big)-\langle\sigma^{X\psi_{\!+}\psi_{\!-} h_i}_{\vmol}\rangle\Big(n_{X}n_{\psi_{\!+}}-\bar n_{X} \bar n_{\psi_{\!+}}\frac{ n_{\psi_{\!-}}}{\bar n_{\psi_{\!-}}}\Big)    \notag\\
&-\langle\sigma^{X\psi_{\!-}\psi_{\!+} h_i}_{\vmol}\rangle\Big(n_{X}n_{\psi_{\!-}}-\bar n_{X} \bar n_{\psi_{\!-}}\frac{ n_{\psi_{\!+}}}{\bar n_{\psi_{\!+}}}\Big)-\langle\sigma^{Xh_i\psi_{\!+}\psi_{\!-} }_{\vmol}\rangle\bar n_{h_i}\Big(n_{X}-\bar n_{X} \frac{ n_{\psi_{\!+}}n_{\psi_{\!-}}}{\bar n_{\psi_{\!+}}\bar n_{\psi_{\!-}}}\Big)    \notag\\
&-\langle\sigma^{XX\psi_+\psi_+}_{\vmol}\rangle\Big(n_{X}^2- \bar n_{X}^2\frac{n_{\psi_+}^2}{\bar n_{\psi_+}^2}\Big)-\langle\sigma^{XX\psi_-\psi_-}_{\vmol}\rangle\Big(n_{X}^2- \bar n_{X}^2\frac{n_{\psi_-}^2}{\bar n_{\psi_-}^2}\Big)  \notag\\
&+\Gamma_{\psi_{\!+}\to X\psi_{\!-}}\Big(n_{\psi_{\!+}}-  \bar n_{\psi_{\!+}}\frac{n_{X}}{\bar n_{X}}\frac{n_{\psi_{\!-}}}{\bar n_{\psi_{\!-}}}\Big),
\label{boltzmann_X}\\
\frac{dn_{\psi_-}}{dt}=& -3H n_{\psi_-}-\langle\sigma^{\psi_-\psi_-\phi\phi^\p}_{\vmol}\rangle\Big(n_{\psi_-}^2-\bar n_{\psi_-}^2\Big)-\langle\sigma^{\psi_-\psi_+ X h_i}_{\vmol}\rangle\Big(n_{\psi_-}n_{\psi_+}-\bar n_{\psi_-} \bar n_{\psi_+}\frac{ n_{X}}{\bar n_{X}}\Big)    \notag\\
&-\langle\sigma^{X\psi_{\!-}\psi_{\!+} h_i}_{\vmol}\rangle\Big(n_{X}n_{\psi_{\!-}}-\bar n_{X} \bar n_{\psi_{\!-}}\frac{ n_{\psi_{\!+}}}{\bar n_{\psi_{\!+}}}\Big)-\langle\sigma^{\psi_- h_iX\psi_+}_{\vmol}\rangle \bar n_{h_i} \Big(n_{\psi_-}- \bar n_{\psi_-}\frac{n_{\psi_+}}{\bar n_{\psi_+}}\frac{ n_{X}}{\bar n_{X}}\Big)\notag\\
&-\langle\sigma^{\psi_-\psi_-XX}_{\vmol}\rangle\Big(n_{\psi_-}^2-\bar n_{\psi_-}^2 \frac{n_{X}^2}{\bar n_{X}^2}\Big)- \langle\sigma^{\psi_-\psi_-\psi_+\psi_+}_{\vmol}\rangle\Big(n_{\psi_-}^2- \bar n_{\psi_-}^2\frac{n_{\psi_+}^2}{\bar n_{\psi_+}^2}\Big)  \notag\\
&+\Gamma_{\psi_+\to X\psi_-} \Big(n_{\psi_+}- \bar n_{\psi_+}\frac{n_{\psi_-}}{\bar n_{\psi_-}}\frac{ n_{X}}{\bar n_{X}}\Big),
\label{boltzmann_psi_m}\\
\frac{dn_{\psi_+}}{dt}=& -3H n_{\psi_+}-\langle\sigma^{\psi_+\psi_+\phi\phi^\p}_{\vmol}\rangle\Big(n_{\psi_+}^2-\bar n_{\psi_+}^2\Big)-\langle\sigma^{\psi_+\psi_- X h_i}_{\vmol}\rangle\Big(n_{\psi_+}n_{\psi_-}-\bar n_{\psi_+} \bar n_{\psi_-}\frac{ n_{X}}{\bar n_{X}}\Big)    \notag\\
&-\langle\sigma^{X\psi_{\!+}\psi_{\!-} h_i}_{\vmol}\rangle\Big(n_{X}n_{\psi_{\!+}}-\bar n_{X} \bar n_{\psi_{\!+}}\frac{ n_{\psi_{\!-}}}{\bar n_{\psi_{\!-}}}\Big)-\langle\sigma^{\psi_+ h_iX\psi_-}_{\vmol}\rangle \bar n_{h_i} \Big(n_{\psi_+}- \bar n_{\psi_+}\frac{n_{\psi_-}}{\bar n_{\psi_-}}\frac{ n_{X}}{\bar n_{X}}\Big)\notag\\
&-\langle\sigma^{\psi_+\psi_+XX}_{\vmol}\rangle\Big(n_{\psi_+}^2-\bar n_{\psi_+}^2 \frac{n_{X}^2}{\bar n_{X}^2}\Big)- \langle\sigma^{\psi_+\psi_+\psi_-\psi_-}_{\vmol}\rangle\Big(n_{\psi_+}^2- \bar n_{\psi_+}^2\frac{n_{\psi_-}^2}{\bar n_{\psi_-}^2}\Big)  \notag\\
& -\Gamma_{\psi_+\to X\psi_-} \Big(n_{\psi_+}- \bar n_{\psi_+}\frac{n_{\psi_-}}{\bar n_{\psi_-}}\frac{ n_{X}}{\bar n_{X}}\Big),	\label{boltzmann_psi_p}
\end{align}
where $\langle\sigma^{ijkl}_{\vmol}\rangle\equiv \langle\sigma^{ijkl}{\vmol}\rangle$ is the thermally averaged cross-section for the process $ij\to kl$ as defined in Eq.~\eqref{sigmaV}. Above $h_i=h_1,h_2$ and $\phi\phi^\p$ denote all the allowed SM particles including $h_1,h_2$. In the above Boltzmann equations~\eqref{boltzmann_X}-\eqref{boltzmann_psi_p}, the first term $3H n_{i}$ is the usual term in an expanding universe with Hubble parameter $H$. The second term in Eqs.~\eqref{boltzmann_X}-\eqref{boltzmann_psi_p} is the standard annihilation term for each dark particle corresponding to Feynman diagrams Fig.~\ref{fig:annihilation}, whereas, the third, fourth and fifth terms are capturing the effects of semi-annihilations shown in Feynman diagrams Fig.~\ref{fig:semi_annihilation}, and the sixth and seventh terms are conversion processes within the dark sector shown in Fig.~\ref{fig:conversion}. Note the last term in these Boltzmann equations is for the case considered in Sec.~\ref{A vector and a Majorana fermion as dark matter} when $\psi_+$ is unstable and it decays to stable particles $X_\mu$ and $\psi_-$. However, for the case discussed in Sec.~\ref{Two Majorana fermions as dark matter} where $X_\mu$ is unstable and it decays to stable particles $\psi_+$ and $\psi_-$, we replace $\psi_+\leftrightarrow X$ and change signs of the last terms in Eqs.~\eqref{boltzmann_X}-\eqref{boltzmann_psi_p}. Moreover, for the case studied in Sec.~\ref{A vector and two Majorana fermions as dark matter} when all three dark particles $X_{\mu},\psi_\pm$ are stable then the last terms in the above Boltzmann equations are zero. The details of the derivation of above collision terms are presented in~\ref{Boltzmann equations for semi-annihilating dark matter}.

After solving the Boltzmann equations we calculate the present relic density of the dark species as, 
\beq
\Omega_i h^2=\frac{h^2 s_0}{\rho_{\text{cr}}}m_iY_i=2.742\times 10^{8}\left(\frac{m_i}{\text{GeV}}\right)Y_i,
\eeq
where $s_0$ is the total entropy density today, $\rho_{\text{cr}}$ is the critical density, $m_i$ is the mass of the dark particle and $Y_i$ is the yield of the dark particle today. 
Total dark matter relic density is a sum of the individual relics, i.e.
\beq
\Omega_{\text{tot}} h^2=\sum_i \Omega_i h^2.
\eeq
The total relic density $\Omega_{\rm tot} h^2$ is compared to the observed dark matter relic density $\Omega_{\text{obs}} h^2\!=\!0.1197 \pm0.0022$ by the PLANCK satellite \cite{Ade:2015xua}. 

The simple form of the derived Boltzmann equations relies on the assumption of kinetic equilibrium between visible and hidden sectors which is maintained by frequent scatterings of dark matter species on relativistic SM states. Effects of kinetic decoupling are negligible in the calculation of relic densities only if its temperature $T_{kd}$ is smaller or comparable to the chemical decoupling temperature, $T_{cd}$, at which DM annihilation is no longer effective. $T_{kd}$ can be estimated by comparing the scattering rates $\Gamma_s(T)$ for the processes ${\rm DM~SM}\rightarrow {\rm DM~SM}$ with the Hubble rate $H(T)$~\cite{Hofmann:2001bi}.
\beq
\Gamma_s(T)=n_r \langle \sigma_s v \rangle \frac{T}{m_{\rm DM}},
\eeq
where $n_r$ is the equilibrium density of relativistic states in the thermal bath, $\langle \sigma_s v \rangle$ is the thermally averaged scattering cross-section and $T/m_{\rm DM}$ describes momentum transfer at each collision. Note $m_{\rm DM}$ is the mass of the dominant DM component, so either $m_X$ or $m_\pm$. Considering scatterings of $\psi_\pm$ and $X$ on SM quarks and leptons we find using the expansion of the scattering amplitudes~\cite{Bringmann:2006mu} that
\beq
T_{kd}\simeq \frac{1.8\times 10^{-8}\;{\rm GeV}^{-3/2}}{\gx\sin\alpha} \frac{\xi\sqrt{m_X} m^2_{h_1} m^2_{h_2}}{|m^2_{h_1}-m^2_{h_2}|},
\eeq
where $\xi\!=\!1$ if scattering of $X$ dominates and  $\xi\!=\!2\sqrt{m_X m_\pm}/(m_+-m_-)$ for the $\psi_\pm$ domination. Choosing typical values that are used in further discussions $\gx\sim \sin\alpha\sim 0.1$ and $m_{\rm DM} \sim {\cal O}(100)\gev$, we obtain $T_{kd}\lesssim 0.5$~GeV which is well below the temperature of the chemical decoupling in this case ($T_{cd}\simeq m_{\rm DM}/20$). It becomes comparable to $T_{cd}$ only for degenerate Higgs masses, but we checked that for the parameters we adopt always $T_{kd}<T_{cd}$.

Even without solving the above coupled set of Boltzmann equations \eqref{boltzmann_X}--\eqref{boltzmann_psi_p} some generic observations could be made:
\bit\itemsep0em
\item Conversions are present even in the absence of the dark sector self-interactions, an existence of a mediator is the only requirement. On the other hand, the existence of semi-annihilations and decays of dark particle depend on the presence a vertex with three dark states, which have different transformation rules under the dark symmetry (such that a singlet could be formed), in our model such an interaction is in Eq.~\eqref{dark_vertex}.
\item When, for a given dark matter species, a standard annihilation channel is suppressed then its abundance might be very sensitive to the presence of other ingredients of the dark segment. In this case semi-annihilation plays a major role, e.g. if $\psi_-\psi_- \to h_ih_i$ (or any SM states) is suppressed then $\psi_-$ can still disappear, for instance, through $\psi_- \psi_+ \to X_\mu h_i$ followed by unsuppressed annihilation of pairs of $X_\mu$, see also~\cite{DEramo:2010keq}. In other words, $X_\mu$ can work as a catalyzer that enables disappearance of $\psi_-$. In this case, it is possible that the presence of other ($\psi_+$ and/or $X_\mu$) dark components might be crucial for the determination of the asymptotic abundance of the major DM element. Also, decays within the dark sector may play a relevant role in the determination of the final abundance. 
\item Standard WIMPs decouple from thermal equilibrium at $m/T \!\sim\!20\!-\!25$, which implies that the heavy states decouple earlier (large $T$). However, in the multi-component scenario, it might be possible that the decoupling of a heavier dark component is delayed so that it happens later than that of a lighter one. The effect is again a consequence of interactions with remaining dark matter states.
\eit
In particular, as a proof of principle, below we consider two interesting possibilities in our specific vector-fermion DM scenario:
\bit\itemsep0em
\item[(A)] \underline{$\yx\ll 1$ ($m_+\simeq m_-$)}: 
Small $\yx$ implies suppressed $\psi_\pm\psi_\pm$ annihilation, so $\psi_\pm$ dominates the dark matter abundance. Since the annihilation is slow therefore $Y_{\psi_\pm}$ is controlled by semi-annihilation which is sensitive to
$\gx$ and to the presence of other dark components. In order to have semi-annihilation controlled exclusively by $\gx$ one should assume $m_++m_- > m_X+m_{h_2}$. 
\item[(B)] \underline{$\yx\gg 1$ ($m_+\gg m_-$)}:  
In this case, one expects fast $\psi_\pm\psi_\pm$ annihilation and so that $X_\mu$ may dominate the dark matter abundance. If in addition $\sin\!\alpha \ll 1$ then $XX$ annihilation 
would be suppressed so that $Y_X$ shall be controlled by semi-annihilation and conversion processes which are sensitive to the gauge coupling $\gx$ and Yukawa coupling $\yx$. In both cases, $X_\mu$ would be effectively replaced by $\psi_\pm$, which then would disappear through enhanced standard annihilation.  
\eit
It is worth to emphasize the importance of the gauge coupling $\gx$ between all the dark components $X_\mu$, $\psi_+$ and $\psi_-$. This is the most relevant coupling which determines interesting aspects of dynamics of dark matter density evolution.
If the gauge coupling $\gx$ was very small then the model would reduce to a simple sum of two non-interacting components ($\psi_+,\psi_-$), i.e. a rather uninteresting scenario. Note that $\gx$ is a consequence of the presence of the $(-,-)$ state and, as illustrated by our vector-fermion model, existence of this state is quite natural. Note that if the $(-,-)$ state would have been absent then only two fermion dark components ($\psi_+,\psi_-$) would have been allowed by the stabilizing symmetries. However, in this case, only annihilation and conversion processes -- without semi-annihilations and decays -- would have been allowed. Again not a very appealing scenario. 

\subsection{Multi-component cases}
\label{Multi-component cases}

In the following subsections we consider various interesting setups with two or three dark particles. 
Matrix elements squared needed for collision terms in the Boltzmann equations \eqref{boltzmann_X}--\eqref{boltzmann_psi_p} are computed by employing the {\tt CalcHEP}~\cite{Belyaev:2012qa}, whereas for thermal averaging and solutions of the Boltzmann equations, we adopt our dedicated {\tt C++} code\,\footnote{Note that the presence of three stable dark matter components is quite generic in models with two interacting stable states. In this case, even the 2-component version of \texttt{micrOMEGAs}~\cite{Belanger:2014vza} is not applicable as the code assumes there are at most two DM sectors within which particles remain in equilibrium. 
Therefore for the case of 3-component DM, we adopt our dedicated code which employs the full set of three Boltzmann equations.}.

%
\subsubsection{2CDM: a vector and a Majorana fermion as dark matter}
\label{A vector and a Majorana fermion as dark matter}
\begin{figure*}[t]
\centering
\includegraphics[width=0.5\textwidth]{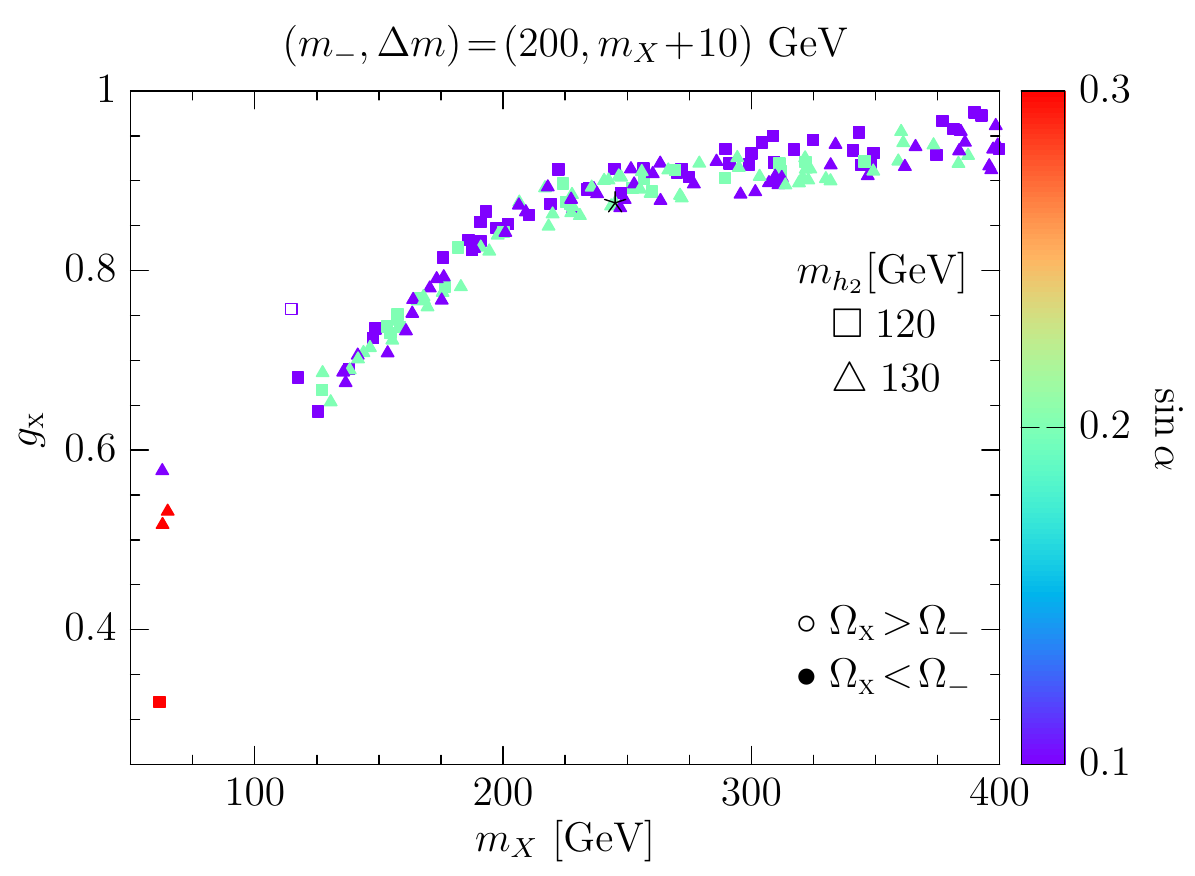}\hspace{-5pt}
\includegraphics[width=0.5\textwidth]{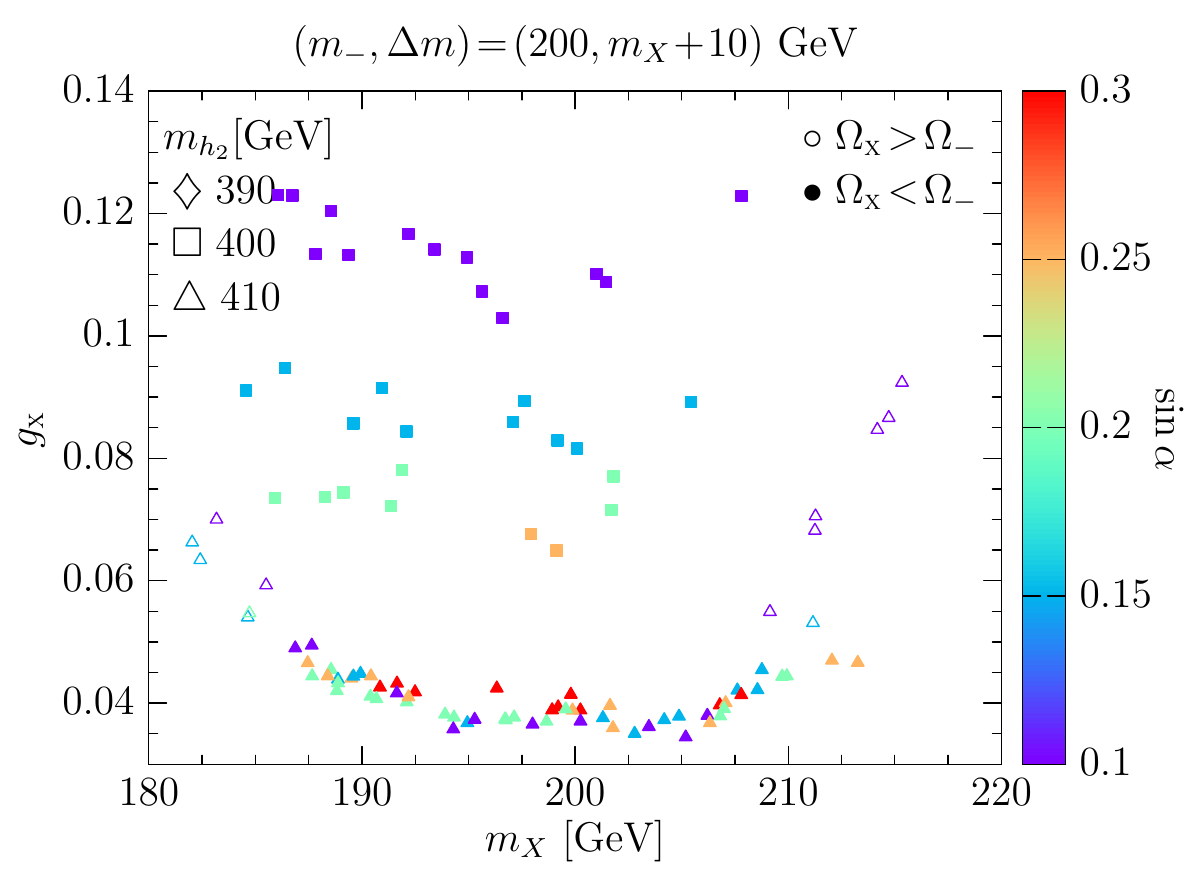}
\includegraphics[width=0.5\textwidth]{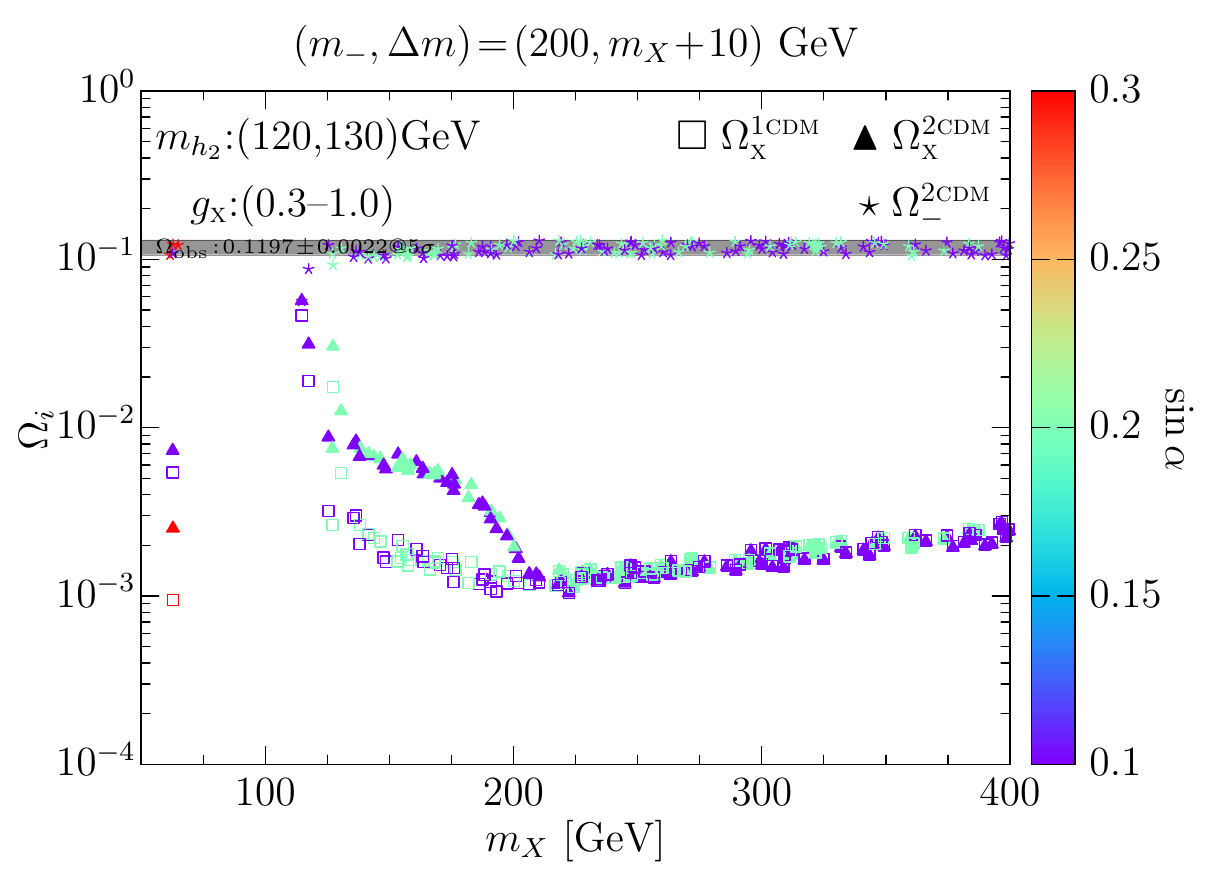}\hspace{-5pt}
\includegraphics[width=0.5\textwidth]{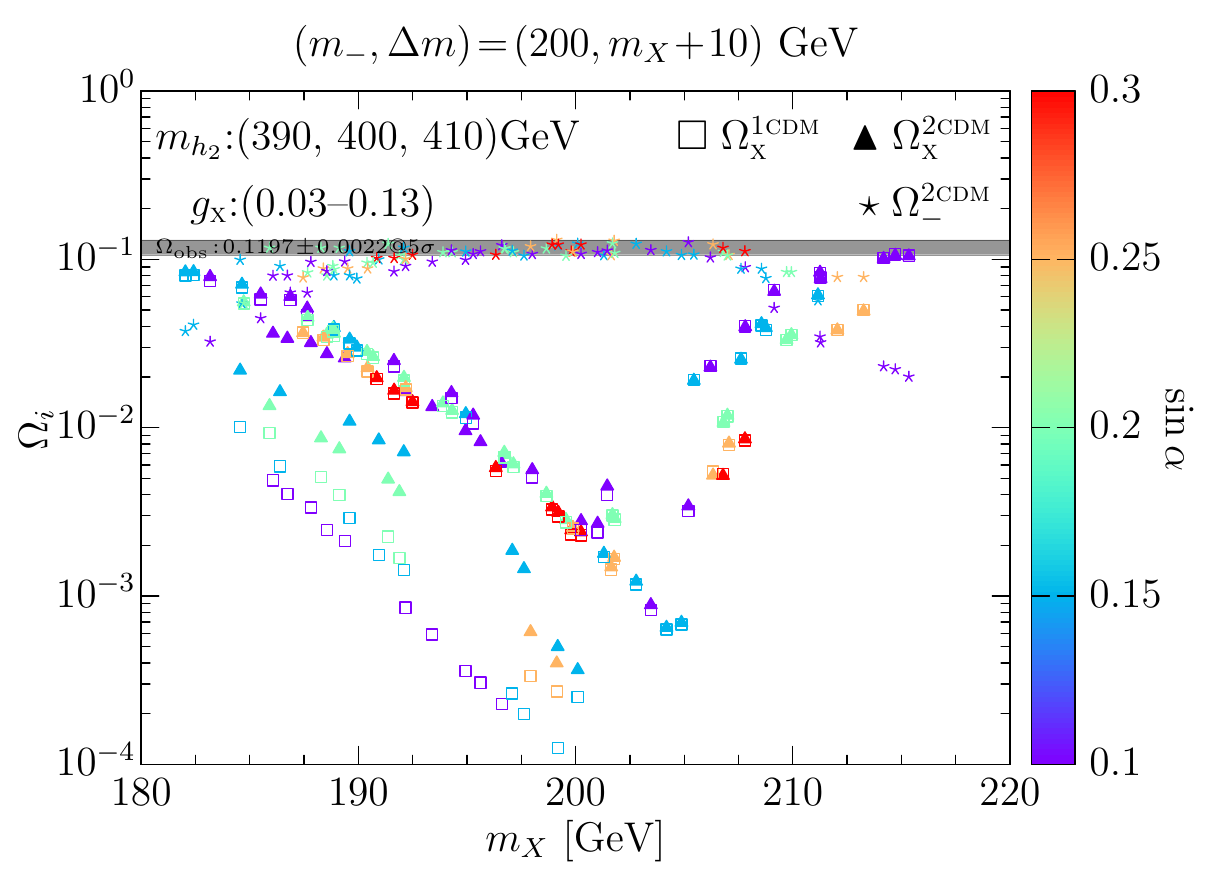}
\caption{Upper panels: The plots show results of scans over the parameter space of a 2CDM case, where $X_\mu$ and $\psi_-$ are stable. All the points  satisfy the correct relic density observed by PLANCK at $5\sigma$ and the recent direct detection experimental bound from LUX2016 at $2\sigma$.
Lower panels:
The plots show results of scans over the parameter space of a 1CDM (assuming the fermions $\psi_\pm$ are decoupled from the dark sector and hence $X_\mu$ is the only dark matter particle) and 2CDM (where $X_\mu$ and $\psi_-$ are stable) cases, respectively. In the 2CDM case, all the points are taken from the corresponding upper panel plots. Note that for the 1CDM case, in most of the parameter space, the single dark matter $X_\mu$ is under-abundant, whereas in the 2CDM the presence of second component $\psi_-$ provides the remaining relic density.
The point denoted by the black star $\star$ located in the upper-left panel at $m_X=245$~GeV corresponds to the same parameters as those adopted in the middle panel of Fig.~\ref{fig:strategy_b}.
}
\label{fig:scanvfdm_Xp}
\end{figure*}

In this subsection we show results for the scenario with $m_+>m_-+m_X$, so that $\psi_+$ is unstable and can decay into lighter Majorana fermion $\psi_-$ and the vector boson $X_\mu$.
Fig.~\ref{fig:scanvfdm_Xp} shows results of a scan over $\sin\!\alpha, \gx, m_X$ for fixed $m_{h_2}, m_-$ 
and $\Delta m=(m_X+10)\gev$. 
All the points satisfy the correct relic density (for the total abundance) observed by PLANCK at $5\sigma$ and the recent direct detection experimental bound from LUX2016 at $2\sigma$. Note that, since the Yukawa coupling to dark fermions is proportional to $\Delta m/v_X \!=\! (\Delta m/m_X) g_X$ therefore for fixed $\Delta m$ and $m_X$ the annihilation cross-sections for the both $\psi_-$ and $X$ are proportional to $(g_X \sin 2\alpha)^2$ therefore, if the remaining parameters are fixed, the requirement of correct DM abundance determines $g_X \sin 2\alpha$. Then the mass of the DM component decides which component contributes more to the $\Omega_{DM}$.
The left panels are for close $h_1$ and $h_2$ masses, $m_{h_2}=120$ and $m_{h_2}=130\gev$ which allows for cancellation between $h_1$ and $h_2$ contributions to the direct detection cross-section on nuclei and therefore for consistency with LUX2016 data even for coupling constants that are not so small ($\gx\!=\!0.3-1.0$ with majority of points located above $\gx\!\sim\! 0.3$). Note that also $\sin\alpha$ does not need to be particularly small due to efficient cancellation between $h_1$ and $h_2$ contributions to the direct detection.

The right panels correspond to parameter points such that around $m_X\!=\!200\gev$ there is a resonant enhancement of $XX$ annihilation through the $h_2$ s-channel exchange. Of course, there is also a non-resonant contribution from $\psi_-\psi_-$ annihilation. In the case of $h_2$ resonance, in order to satisfy the relic abundance constraint, the relevant coupling constants must be small, i.e. $g_X \sin\alpha \ll 1$, it turns out that for the region of $\sin\alpha$ considered here the  gauge coupling constants  must be in the range $\gx\!\sim\!0.03-0.13$.

Lower panels in Fig.~\ref{fig:scanvfdm_Xp} show the abundances of both components separately, $\Omega_{X}^{\textsc{2cdm}}$ and $\Omega_{-}^{\textsc{2cdm}}$, for our model and corresponding $\Omega_X^{\textsc{1cdm}}$ calculated in a 1-component VDM assuming the Majorana fermions are decoupled (which is achieved by $m_D\to\infty$). The abundance and direct detection cross-section in the VDM model were calculated for the same parameters as those adopted in the 2-component model, just truncated to $m_{h_2}, m_X, \sin\!\alpha, \gx$. It is worth to focus at $X$ masses between $110$ and $200\gev$ in the lower-left panel, where for each given $m_X$ there is a clear shift upwards of the $X$ abundance in the 2CDM scenario (filled triangles are above empty boxes). This is a nice illustration how the presence of the other components of the dark sector may influence the abundance of $X$. Similarly, in the lower-right plot, an analogous shift is observed for $m_X \lsim 200\gev$, this shift is enhanced by small $\sin\!\alpha$, e.g. for $\sin\!\alpha\sim 0.1$ the $X$ abundance receives an extra factor $\sim 10^2$. Usually, the presence of other constituents of the dark sector implies both an upward shift of $\Omega_X$, when compared to the 1CDM, and provides a necessary extra contribution by $\Omega_-^{\textsc{2cdm}}$ in order to satisfy the abundance constraint.

\begin{figure*}[t]
\centering
\begin{minipage}[h]{0.33\textwidth}
\includegraphics[width=\textwidth]{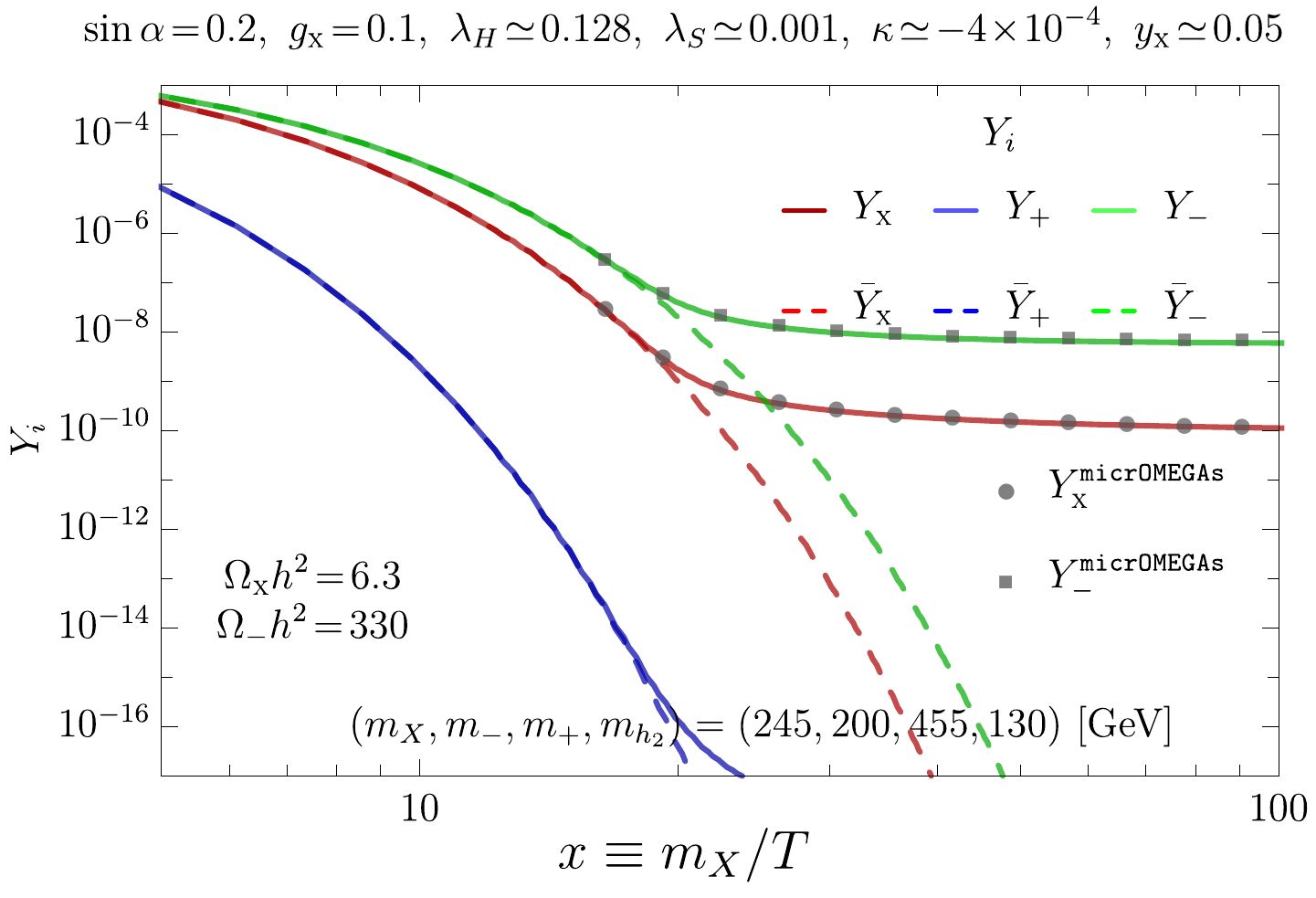}
\resizebox{\textwidth}{!}{%
\begin{tabular}{|c|c|c|c|}
\hline
process & $a_N$ & $a_{N+1}$ & $N$ \\
\hline
$XX \rightarrow \text{SM}$ & $1.2\cdot 10^{-2}$ & $-2.9\cdot 10^{-2}$ & $0$ \\
$\psi_+\psi_+ \rightarrow \text{SM}$ & $2.3\cdot 10^{-4}$ & $-5.7\cdot 10^{-4}$ & $1$ \\
$\psi_-\psi_- \rightarrow \text{SM}$ & $6.8\cdot 10^{-3}$ & $-4.3\cdot 10^{-2}$ & $1$ \\
\hline
$\psi_+\psi_+ \rightarrow XX$ & $6.8\cdot 10^{-4}$ & $1.3\cdot 10^{-3}$ & $0$ \\
$XX \rightarrow \psi_-\psi_-$ & $6.4\cdot 10^{-5}$ & $1.3\cdot 10^{-4}$ & $0$ \\
$\psi_+\psi_+ \rightarrow \psi_-\psi_-$ & $3.9\cdot 10^{-3}$ & $1.6\cdot 10^{-3}$ & $0$ \\
\hline
$\psi_+\psi_- \rightarrow Xh_1$ & $3.9\cdot 10^{-6}$ & $1.7\cdot 10^{-5}$ & $0$ \\
$\psi_+\psi_- \rightarrow Xh_2$ & $9.3\cdot 10^{-5}$ & $4.3\cdot 10^{-4}$ & $0$ \\
$\psi_+h_1 \rightarrow X\psi_-$ & $2.2\cdot 10^{-4}$ & $3.7\cdot 10^{-3}$ & $0$ \\
$\psi_+h_2 \rightarrow X\psi_-$ & $5.4\cdot 10^{-3}$ & $8.3\cdot 10^{-2}$ & $0$ \\
$X\psi_+ \rightarrow \psi_-h_1$ & $2.9\cdot 10^{-4}$ & $-2.7\cdot 10^{-4}$ & $0$ \\
$X\psi_+ \rightarrow \psi_-h_2$ & $6.9\cdot 10^{-3}$ & $-6.6\cdot 10^{-3}$ & $0$ \\
\hline
$\psi_+ \rightarrow X\psi_-$ & \multicolumn{3}{|c|}{$1.9\cdot 10^{-3}$} \\
\hline
\end{tabular}}
\end{minipage}
\!\begin{minipage}[h]{0.33\textwidth}
\includegraphics[width=\textwidth]{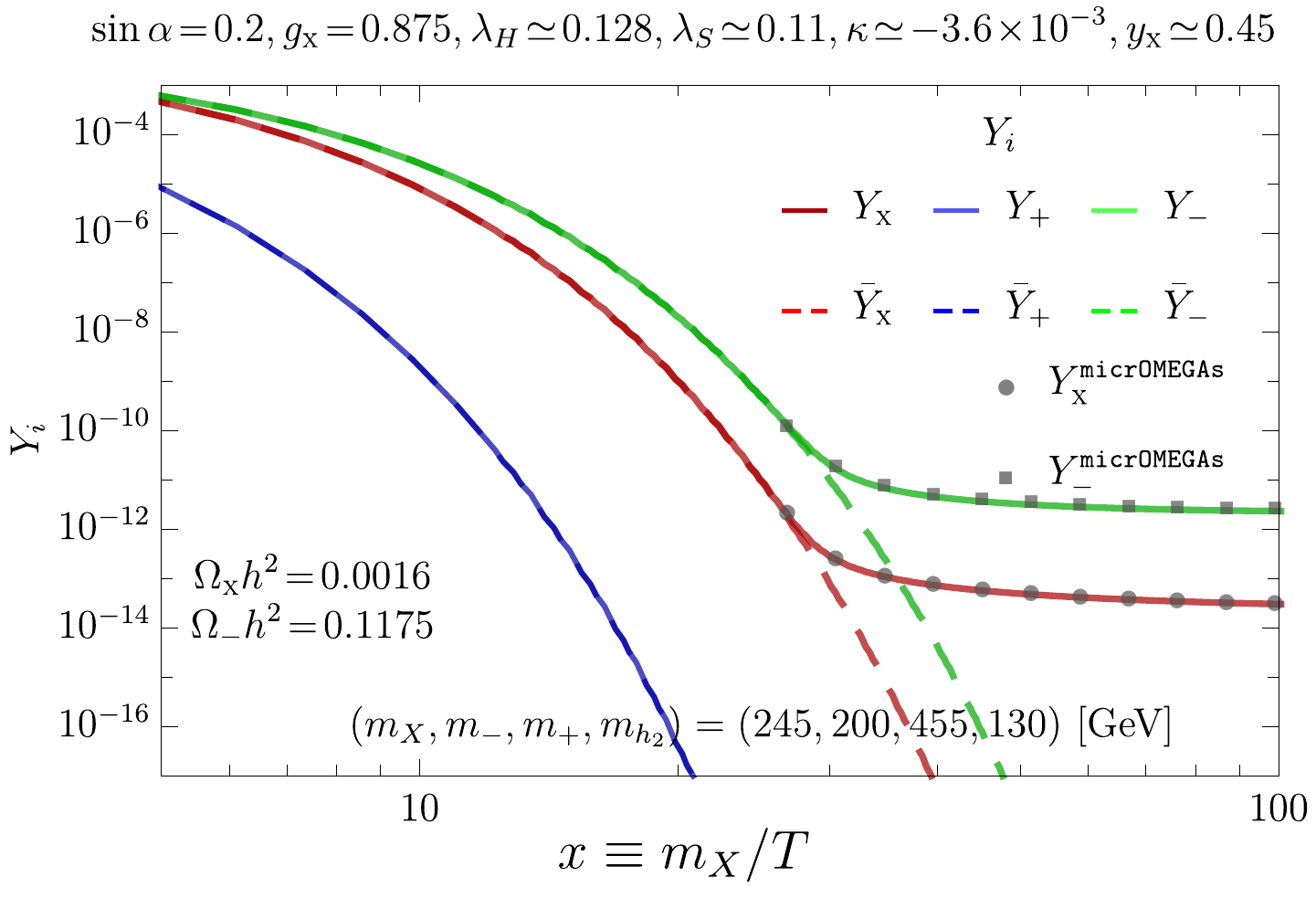}
\resizebox{\textwidth}{!}{%
\begin{tabular}{|c|c|c|c|}
\hline
process & $a_N$ & $a_{N+1}$ & $N$ \\
\hline
$XX \rightarrow \text{SM}$ & $7.2\cdot 10^{1}$ & $-1.69\cdot 10^{2}$ & $0$ \\
$\psi_+\psi_+ \rightarrow \text{SM}$ & $1.36\cdot 10^{0}$ & $-3.36\cdot 10^{0}$ & $1$ \\
$\psi_-\psi_- \rightarrow \text{SM}$ & $4\cdot 10^{1}$ & $-2.55\cdot 10^{2}$ & $1$ \\
\hline
$\psi_+\psi_+ \rightarrow XX$ & $3.97\cdot 10^{0}$ & $7.59\cdot 10^{0}$ & $0$ \\
$XX \rightarrow \psi_-\psi_-$ & $3.7\cdot 10^{-1}$ & $7.6\cdot 10^{-1}$ & $0$ \\
$\psi_+\psi_+ \rightarrow \psi_-\psi_-$ & $2.27\cdot 10^{1}$ & $9.15\cdot 10^{0}$ & $0$ \\
\hline
$\psi_+\psi_- \rightarrow Xh_1$ & $2.3\cdot 10^{-2}$ & $1\cdot 10^{-1}$ & $0$ \\
$\psi_+\psi_- \rightarrow Xh_2$ & $5.5\cdot 10^{-1}$ & $2.53\cdot 10^{0}$ & $0$ \\
$\psi_+h_1 \rightarrow X\psi_-$ & $1.28\cdot 10^{0}$ & $2.16\cdot 10^{1}$ & $0$ \\
$\psi_+h_2 \rightarrow X\psi_-$ & $3.17\cdot 10^{1}$ & $4.85\cdot 10^{2}$ & $0$ \\
$X\psi_+ \rightarrow \psi_-h_1$ & $1.68\cdot 10^{0}$ & $-1.6\cdot 10^{0}$ & $0$ \\
$X\psi_+ \rightarrow \psi_-h_2$ & $4.04\cdot 10^{1}$ & $-3.89\cdot 10^{1}$ & $0$ \\
\hline
$\psi_+ \rightarrow X\psi_-$ & \multicolumn{3}{|c|}{$1.4\cdot 10^{-1}$} \\
\hline
\end{tabular}}
\end{minipage}
\!\begin{minipage}[h]{0.33\textwidth}
\includegraphics[width=\textwidth]{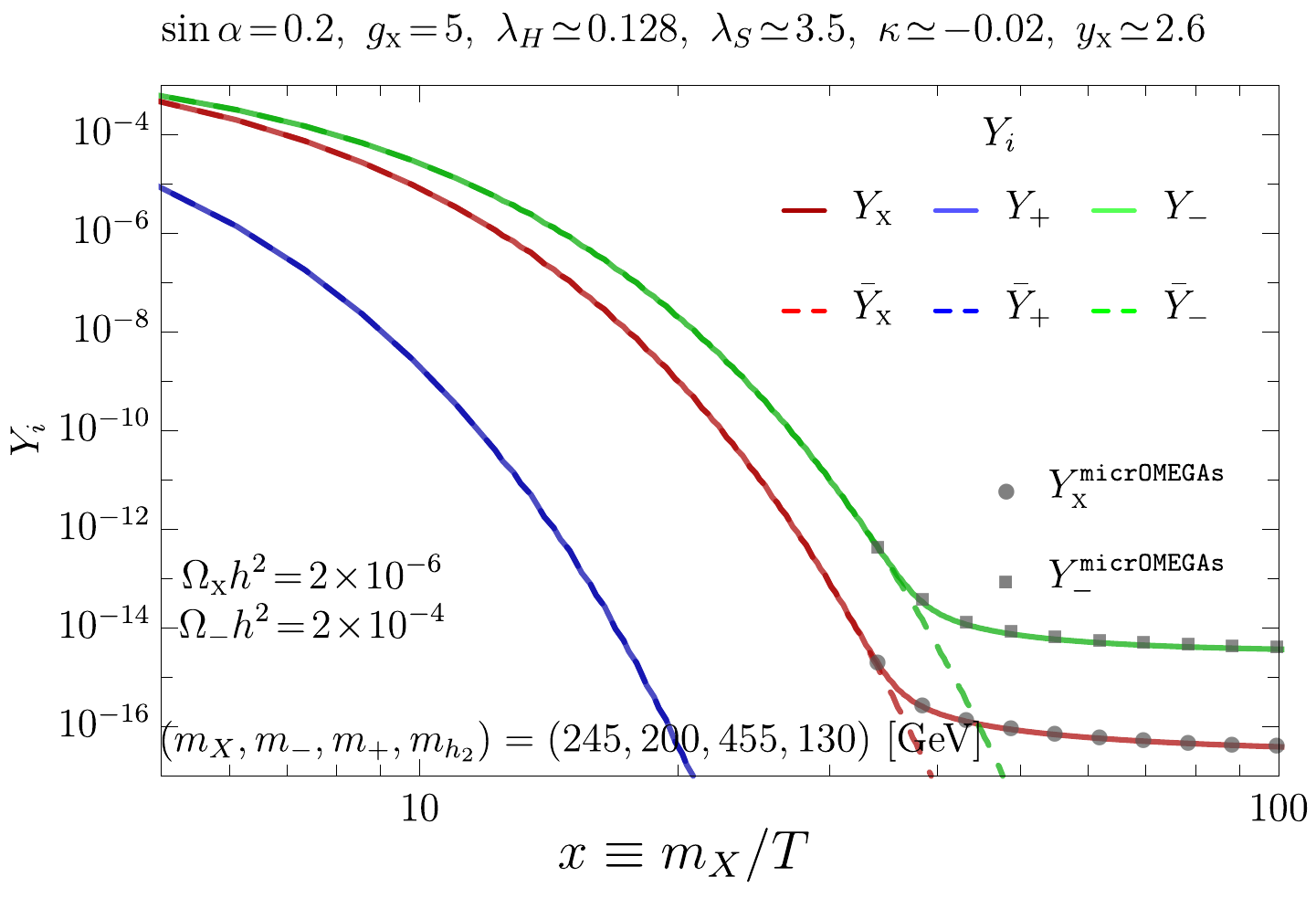}
\resizebox{\textwidth}{!}{%
\begin{tabular}{|c|c|c|c|}
\hline
process & $a_N$ & $a_{N+1}$ & $N$ \\
\hline
$XX \rightarrow \text{SM}$ & $7.68\cdot 10^{4}$ & $-1.78\cdot 10^{5}$ & $0$ \\
$\psi_+\psi_+ \rightarrow \text{SM}$ & $1.45\cdot 10^{3}$ & $-3.58\cdot 10^{3}$ & $1$ \\
$\psi_-\psi_- \rightarrow \text{SM}$ & $4.27\cdot 10^{4}$ & $-2.74\cdot 10^{5}$ & $1$ \\
\hline
$\psi_+\psi_+ \rightarrow XX$ & $4.23\cdot 10^{3}$ & $8.18\cdot 10^{3}$ & $0$ \\
$XX \rightarrow \psi_-\psi_-$ & $3.98\cdot 10^{2}$ & $7.98\cdot 10^{2}$ & $0$ \\
$\psi_+\psi_+ \rightarrow \psi_-\psi_-$ & $2.42\cdot 10^{4}$ & $9.73\cdot 10^{3}$ & $0$ \\
\hline
$\psi_+\psi_- \rightarrow Xh_1$ & $2.44\cdot 10^{1}$ & $1.08\cdot 10^{2}$ & $0$ \\
$\psi_+\psi_- \rightarrow Xh_2$ & $5.84\cdot 10^{2}$ & $2.69\cdot 10^{3}$ & $0$ \\
$\psi_+h_1 \rightarrow X\psi_-$ & $1.37\cdot 10^{3}$ & $2.3\cdot 10^{4}$ & $0$ \\
$\psi_+h_2 \rightarrow X\psi_-$ & $3.37\cdot 10^{4}$ & $5.17\cdot 10^{5}$ & $0$ \\
$X\psi_+ \rightarrow \psi_-h_1$ & $1.8\cdot 10^{3}$ & $-1.71\cdot 10^{3}$ & $0$ \\
$X\psi_+ \rightarrow \psi_-h_2$ & $4.31\cdot 10^{4}$ & $-4.09\cdot 10^{4}$ & $0$ \\
\hline
$\psi_+ \rightarrow X\psi_-$ & \multicolumn{3}{|c|}{$4.71\cdot 10^{0}$} \\
\hline
\end{tabular}}
\end{minipage}
\caption{Solutions of the Boltzmann equations for three sample points from the 2CDM case where $X_\mu$ and $\psi_-$ are stable are shown in the upper panels. These plots show the evolution of the dark matter particle yield $Y_i(x) \equiv n_i/s$ ($n_i$ is the number density, $s$ is the total entropy density and $x$ is defined as $x \equiv m_X/T$) for different species $i=\psi_+,\psi_-$ and $X_\mu$. The colored curves are obtained by making use of a dedicated {\tt C++} code to solve the corresponding Boltzmann equations, whereas the gray points are the corresponding results from the {\tt micrOMEGAs} code. The left, middle and right plots are for the values of parameter $\gx=0.02,0.2$ and $1$, respectively. The values of other model parameters are shown in the legends of plots and the relic density of two dark matter particles $\Omega_{X,-} \; h^2$ is also given in each plot. 
The tables contain first two non-vanishing coefficients of thermally-averaged cross-sections~[pb] expanded in powers of $x^{-1}$, given by $\langle \sigma^{ijkl}\vmol\rangle\! =\! a_Nx^{-N} + a_{N+1}x^{-(N+1)} + \cdots$, and the decay width $\langle \Gamma_{\psi_+\rightarrow X\psi_-}\rangle$~[GeV].
}
\label{fig:strategy_b}
\end{figure*}
In the plots of Fig.~\ref{fig:strategy_b}, and similar figures in the following sections, we show dark matter yields $Y_i(x) \!\equiv\! n_i/s$ ($s$ is the total entropy density and $x$ is defined as $x \!\equiv\! m_X/T$) for different species $i=\psi_+,\psi_-$ and $X_\mu$. Note that we plot bare values of yields, with no extra normalization adopted. Moreover, the tables in Fig.~\ref{fig:strategy_b} and similar figures in the following section contain first two non-vanishing coefficients of thermally-averaged cross-sections~[pb] expanded in powers of $x^{-1}$, given by $\langle \sigma^{ijkl}\vmol\rangle\! =\! a_Nx^{-N} + a_{N+1}x^{-(N+1)} + \cdots$, and the decay width $\langle \Gamma_{\psi_+\rightarrow X\psi_-}\rangle$~[GeV]. 
The plots in Fig.~\ref{fig:strategy_b} illustrate solutions of the Boltzmann equations for three selected sample points. The middle panel shows solutions for parameters that reproduce correct total DM abundance and also satisfy the direct detection LUX limits, so the corresponding point is also present in the scan results shown in the upper-left panel of Fig.~\ref{fig:scanvfdm_Xp} as a black star $\star$. In order to illustrate the relevance of the $\gx$ coupling, the left, middle and right plots are obtained for $\gx=0.02, 0.2$ and $1$, respectively, while other parameters remain unchanged. It is clear that the dependence of the abundance for the major DM component ($\psi_-$) on $\gx$ is very strong. The gray dots and boxes show results obtained using the {\tt micrOMEGAs} code for 2CDM \cite{Belanger:2014vza}. As it is seen in the plots they agree very well with the solid lines which correspond to solutions obtained from our dedicated {\tt C++} code that solves the set of three Boltzmann equations (\ref{boltzmann_X} --\ref{boltzmann_psi_p}). In order to identify the most important processes for a given parameter set, in the tables below the panels in Fig.~\ref{fig:strategy_b} we collect the first two non-vanishing coefficients in the expansion of thermally averaged cross-sections in powers of $x^{-1}$. 

Let's look closer at the middle table of Fig.~\ref{fig:strategy_b}. The cross-sections shown there correspond to the point in the parameter space marked by $\star$ which is located in the upper left panel of Fig.~\ref{fig:scanvfdm_Xp} at $m_X = 245$~GeV. As seen from the middle upper panel of Fig.~\ref{fig:strategy_b} the abundance is dominated by $\psi_-$, and $X$ is a sub-leading component abundance of which is by nearly two orders of magnitude smaller than for $\psi_-$, while the abundance of $\psi_+$ is absolutely negligible. Note that both $\psi_-$ and $X$ decouple from equilibrium roughly at the same temperature, this is the first signal that there must be some correlation between annihilation mechanisms responsible for their disappearance. Since the abundance of $\psi_+$ could be neglected the only relevant processes are $XX \rightarrow \text{SM}$, $\psi_-\psi_- \rightarrow \text{SM}$ and $XX \rightarrow \psi_-\psi_-$. Note that the ratio of cross-sections for the first and the second process is $\sim 1.8$ while their abundances differ by almost two orders of magnitude, therefore the process for additional depletion of $X$ abundance must be $XX \rightarrow \psi_-\psi_-$. This illustrates how sub-leading components may influence the abundance of a dominant component.

\subsubsection{2CDM: two Majorana fermions as dark matter}
\label{Two Majorana fermions as dark matter}
In this subsection we show results for the scenario with $m_X>m_-+m_+$, so $X_\mu$ is unstable and can decay into the Majorana fermions $\psi_-$ and $\psi_+$. 
\begin{figure*}[t]
\centering
\includegraphics[width=0.5\textwidth]{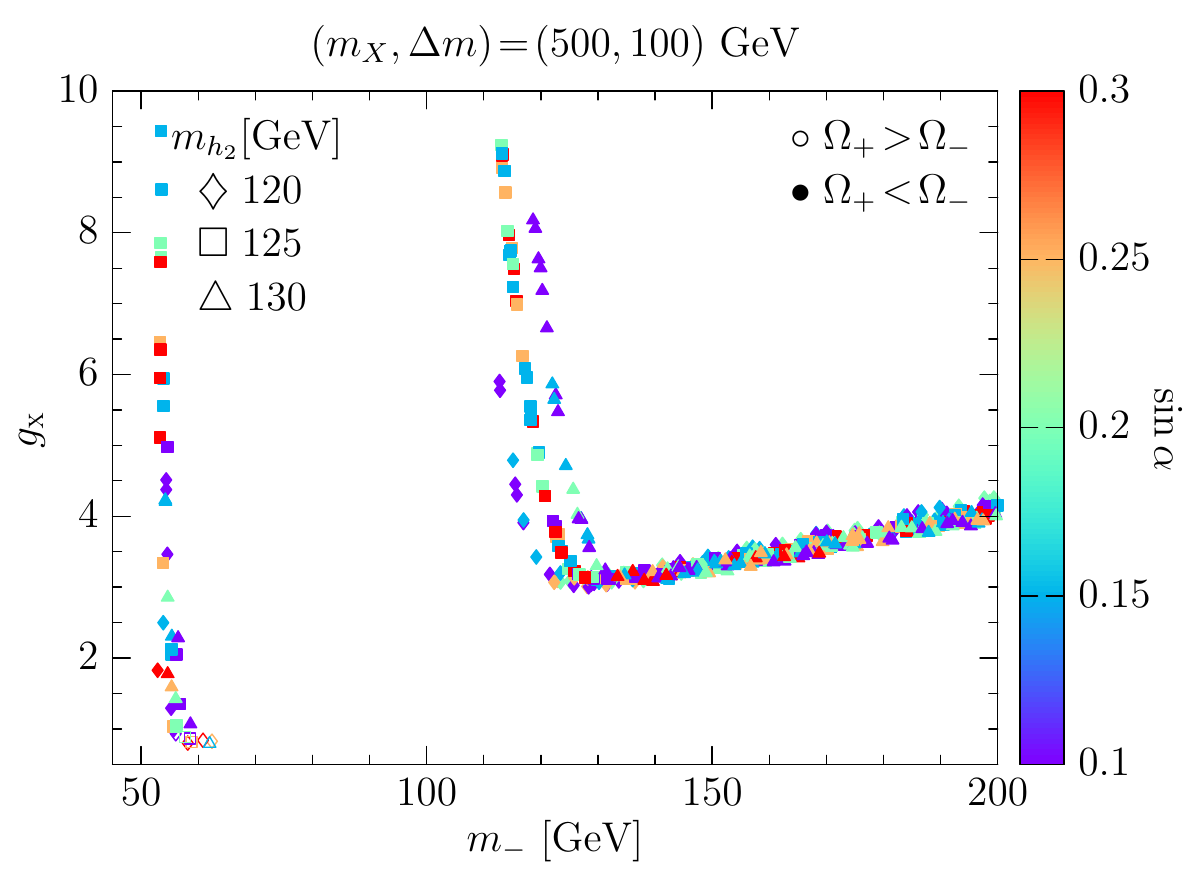}\hspace{-5pt}
\includegraphics[width=0.5\textwidth]{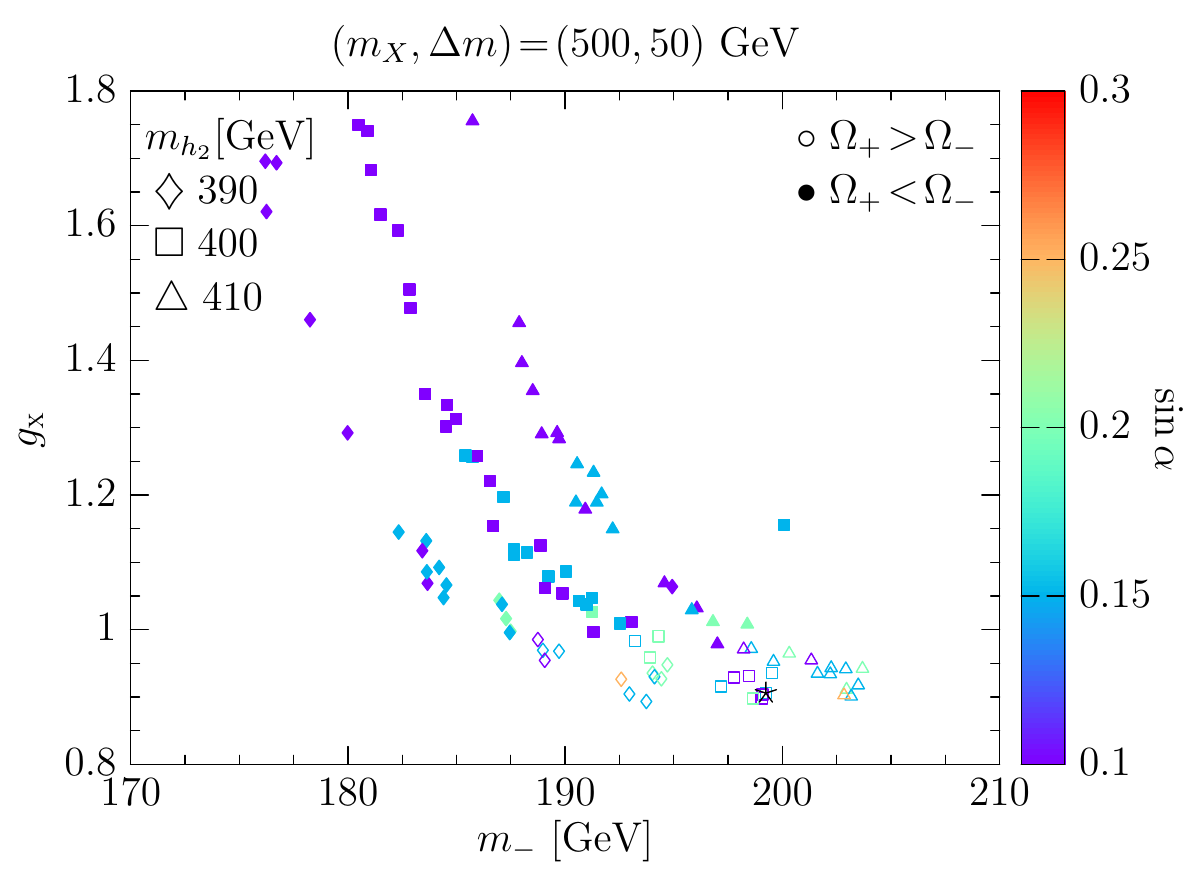}
\caption{The plots show results of scans over the parameter space of a 2CDM case, for which $\psi_+$ and $\psi_-$ are stable. All the points  satisfy the correct relic density observed by PLANCK at $5\sigma$ and the recent direct detection experimental bound from LUX2016 at $2\sigma$.
The point denoted by the star $\star$ located in the right panel at $m_-=199.25$~GeV corresponds to the same parameters as those adopted in the middle panel of Fig.~\ref{fig:degenerate_b}.
}
\label{fig:scan2fdm_pp}
\end{figure*}
\begin{figure*}[t]
\centering
\begin{minipage}[h]{0.33\textwidth}
\includegraphics[width=\textwidth]{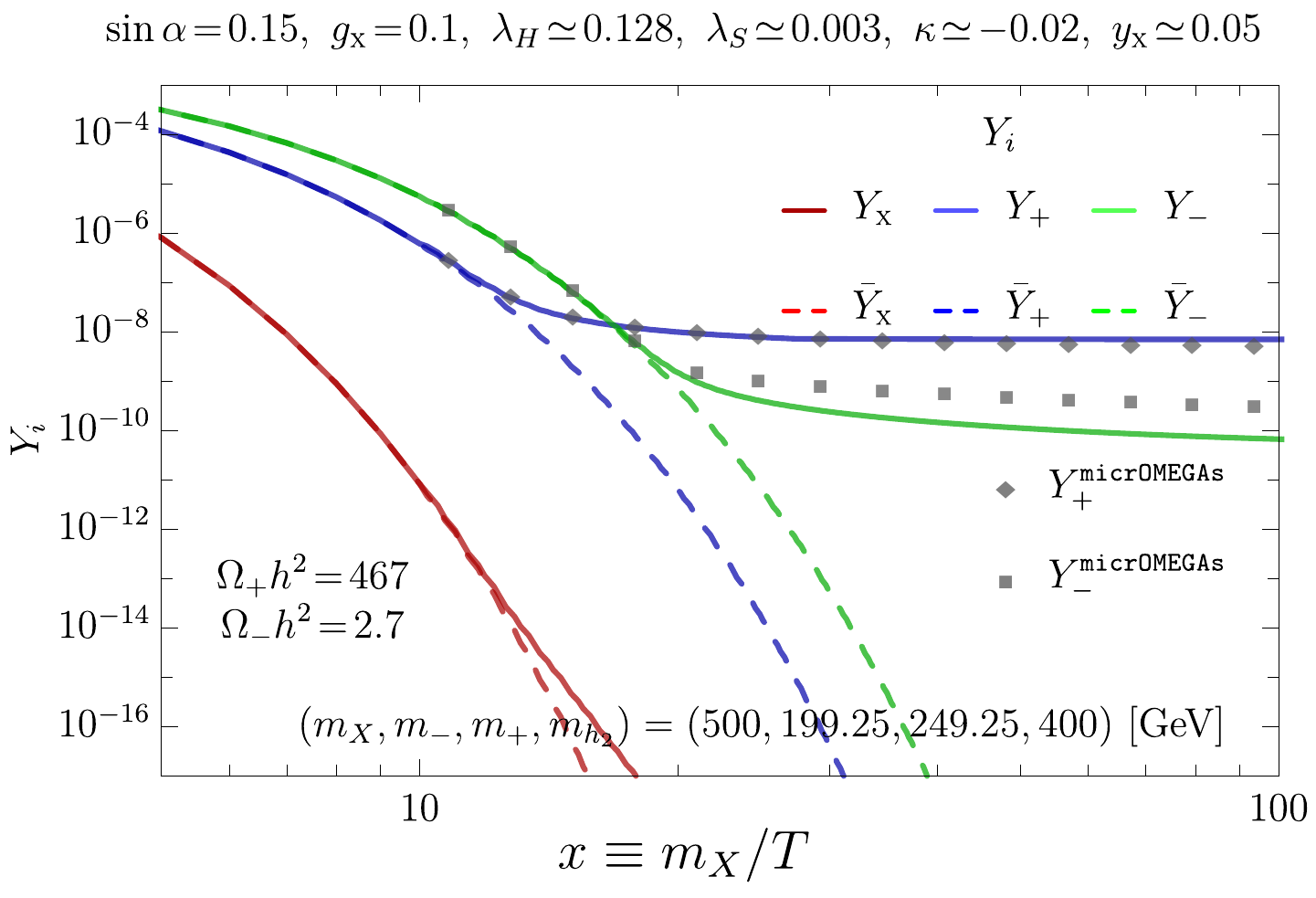}
\resizebox{\textwidth}{!}{%
\begin{tabular}{|c|c|c|c|}
\hline
process & $a_N$ & $a_{N+1}$ & $N$ \\
\hline
$XX \rightarrow \text{SM}$ & $3.2\cdot 10^{-3}$ & $-4.3\cdot 10^{-3}$ & $0$ \\
$\psi_+\psi_+ \rightarrow \text{SM}$ & $5.2\cdot 10^{-4}$ & $-6.6\cdot 10^{-3}$ & $1$ \\
$\psi_-\psi_- \rightarrow \text{SM}$ & $3.91\cdot 10^{0}$ & $2.82\cdot 10^{3}$ & $1$ \\
\hline
$XX \rightarrow \psi_+\psi_+$ & $6.6\cdot 10^{-4}$ & $-1.4\cdot 10^{-3}$ & $0$ \\
$XX \rightarrow \psi_-\psi_-$ & $5.2\cdot 10^{-4}$ & $-1\cdot 10^{-3}$ & $0$ \\
$\psi_+\psi_+ \rightarrow \psi_-\psi_-$ & $1.4\cdot 10^{-4}$ & $5.1\cdot 10^{-4}$ & $0$ \\
\hline
$Xh_1 \rightarrow \psi_+\psi_-$ & $7.9\cdot 10^{-4}$ & $-5.5\cdot 10^{-3}$ & $0$ \\
$Xh_2 \rightarrow \psi_+\psi_-$ & $1.5\cdot 10^{-3}$ & $-3.7\cdot 10^{-3}$ & $0$ \\
$X\psi_- \rightarrow \psi_+h_1$ & $1.1\cdot 10^{-5}$ & $2.6\cdot 10^{-5}$ & $0$ \\
$X\psi_- \rightarrow \psi_+h_2$ & $1.7\cdot 10^{-4}$ & $8.6\cdot 10^{-4}$ & $0$ \\
$X\psi_+ \rightarrow \psi_-h_1$ & $1.3\cdot 10^{-5}$ & $2.1\cdot 10^{-5}$ & $0$ \\
$X\psi_+ \rightarrow \psi_-h_2$ & $2.4\cdot 10^{-4}$ & $1.4\cdot 10^{-3}$ & $0$ \\
\hline
$X \rightarrow \psi_+\psi_-$ & \multicolumn{3}{|c|}{$2\cdot 10^{-2}$} \\
\hline
\end{tabular}}
\end{minipage}
\!\begin{minipage}[h]{0.33\textwidth}
\includegraphics[width=\textwidth]{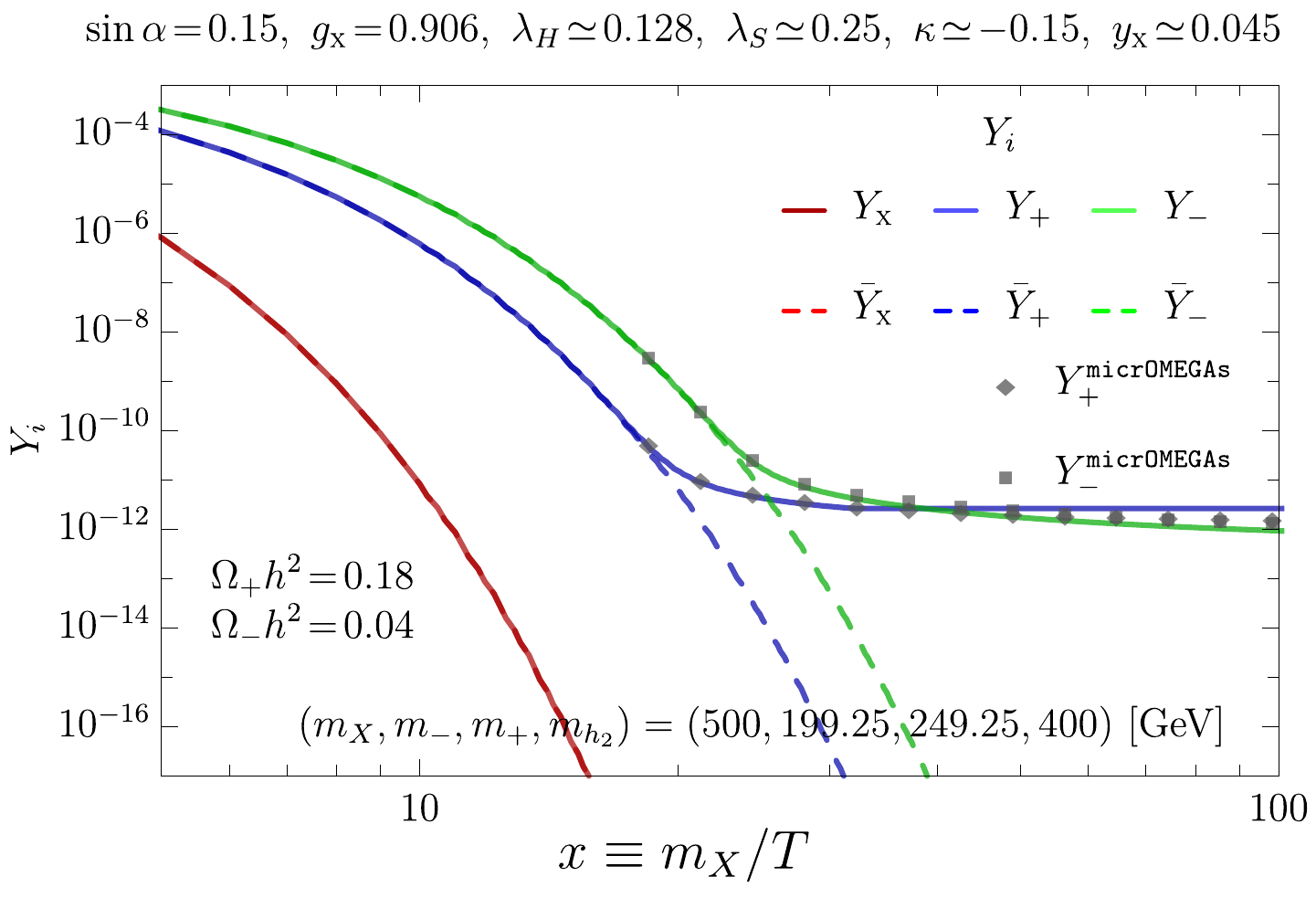}
\resizebox{\textwidth}{!}{%
\begin{tabular}{|c|c|c|c|}
\hline
process & $a_N$ & $a_{N+1}$ & $N$ \\
\hline
$XX \rightarrow \text{SM}$ & $1.38\cdot 10^{1}$ & $-1.41\cdot 10^{1}$ & $0$ \\
$\psi_+\psi_+ \rightarrow \text{SM}$ & $4.1\cdot 10^{-2}$ & $-5.3\cdot 10^{-1}$ & $1$ \\
$\psi_-\psi_- \rightarrow \text{SM}$ & $3.1\cdot 10^{2}$ & $2.24\cdot 10^{5}$ & $1$ \\
\hline
$XX \rightarrow \psi_+\psi_+$ & $4.47\cdot 10^{0}$ & $-9.74\cdot 10^{0}$ & $0$ \\
$XX \rightarrow \psi_-\psi_-$ & $3.54\cdot 10^{0}$ & $-6.93\cdot 10^{0}$ & $0$ \\
$\psi_+\psi_+ \rightarrow \psi_-\psi_-$ & $9.6\cdot 10^{-1}$ & $3.43\cdot 10^{0}$ & $0$ \\
\hline
$Xh_1 \rightarrow \psi_+\psi_-$ & $5.3\cdot 10^{0}$ & $-3.68\cdot 10^{1}$ & $0$ \\
$Xh_2 \rightarrow \psi_+\psi_-$ & $1.04\cdot 10^{1}$ & $-2.47\cdot 10^{1}$ & $0$ \\
$X\psi_- \rightarrow \psi_+h_1$ & $7.2\cdot 10^{-2}$ & $1.7\cdot 10^{-1}$ & $0$ \\
$X\psi_- \rightarrow \psi_+h_2$ & $1.12\cdot 10^{0}$ & $5.82\cdot 10^{0}$ & $0$ \\
$X\psi_+ \rightarrow \psi_-h_1$ & $9\cdot 10^{-2}$ & $1.4\cdot 10^{-1}$ & $0$ \\
$X\psi_+ \rightarrow \psi_-h_2$ & $1.62\cdot 10^{0}$ & $9.33\cdot 10^{0}$ & $0$ \\
\hline
$X \rightarrow \psi_+\psi_-$ & \multicolumn{3}{|c|}{$1.66\cdot 10^{0}$} \\
\hline
\end{tabular}}
\end{minipage}
\!\begin{minipage}[h]{0.33\textwidth}
\includegraphics[width=\textwidth]{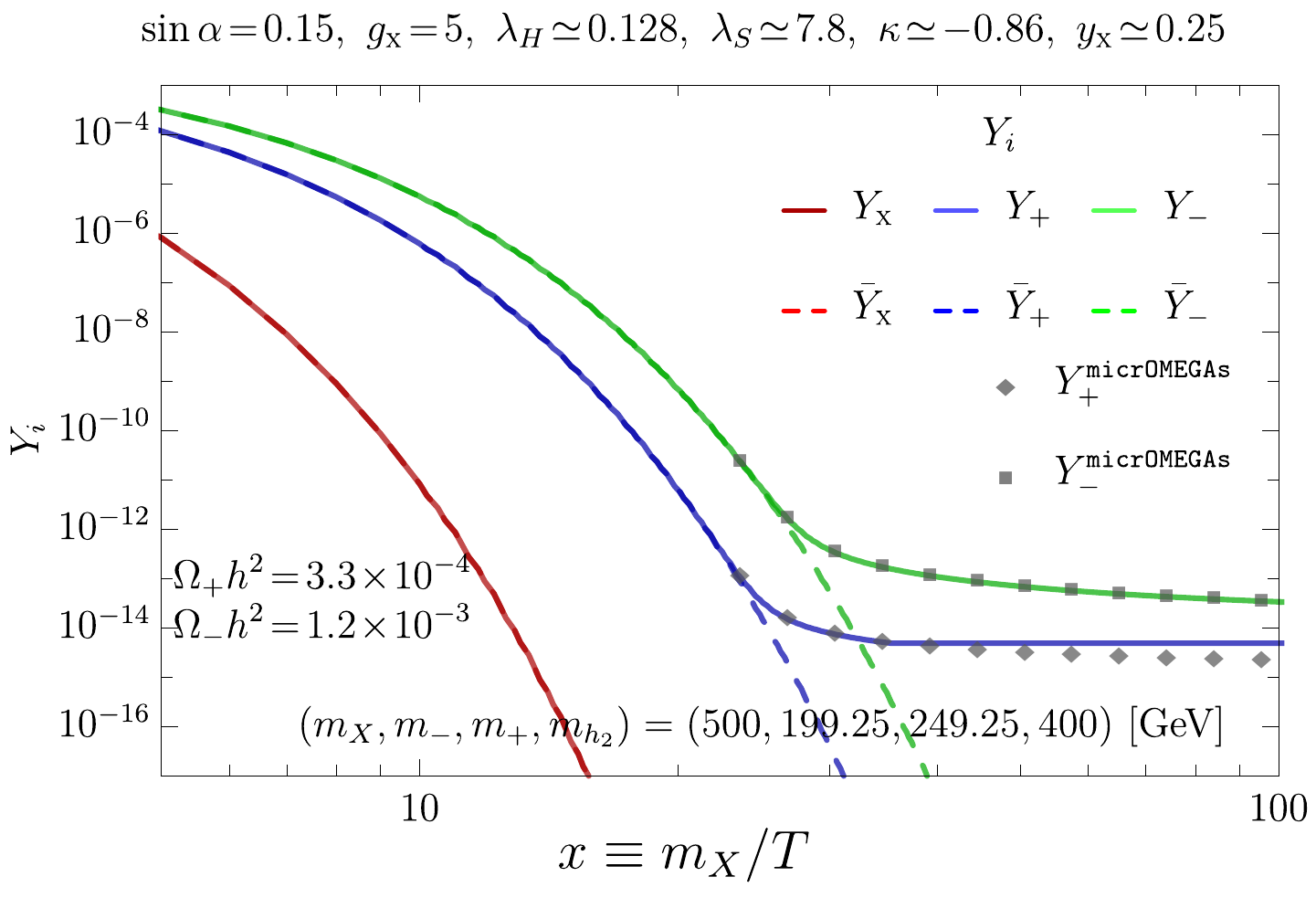}
\resizebox{\textwidth}{!}{%
\begin{tabular}{|c|c|c|c|}
\hline
process & $a_N$ & $a_{N+1}$ & $N$ \\
\hline
$XX \rightarrow \text{SM}$ & $1.27\cdot 10^{4}$ & $-1.27\cdot 10^{4}$ & $0$ \\
$\psi_+\psi_+ \rightarrow \text{SM}$ & $1.1\cdot 10^{0}$ & $-1.4\cdot 10^{1}$ & $1$ \\
$\psi_-\psi_- \rightarrow \text{SM}$ & $8.07\cdot 10^{3}$ & $5.83\cdot 10^{6}$ & $1$ \\
\hline
$XX \rightarrow \psi_+\psi_+$ & $4.15\cdot 10^{3}$ & $-8.96\cdot 10^{3}$ & $0$ \\
$XX \rightarrow \psi_-\psi_-$ & $3.28\cdot 10^{3}$ & $-6.44\cdot 10^{3}$ & $0$ \\
$\psi_+\psi_+ \rightarrow \psi_-\psi_-$ & $8.91\cdot 10^{2}$ & $3.19\cdot 10^{3}$ & $0$ \\
\hline
$Xh_1 \rightarrow \psi_+\psi_-$ & $4.92\cdot 10^{3}$ & $-3.41\cdot 10^{4}$ & $0$ \\
$Xh_2 \rightarrow \psi_+\psi_-$ & $9.65\cdot 10^{3}$ & $-2.3\cdot 10^{4}$ & $0$ \\
$X\psi_- \rightarrow \psi_+h_1$ & $6.7\cdot 10^{1}$ & $1.62\cdot 10^{2}$ & $0$ \\
$X\psi_- \rightarrow \psi_+h_2$ & $1.04\cdot 10^{3}$ & $5.39\cdot 10^{3}$ & $0$ \\
$X\psi_+ \rightarrow \psi_-h_1$ & $8.3\cdot 10^{1}$ & $1.34\cdot 10^{2}$ & $0$ \\
$X\psi_+ \rightarrow \psi_-h_2$ & $1.5\cdot 10^{3}$ & $8.66\cdot 10^{3}$ & $0$ \\
\hline
$X \rightarrow \psi_+\psi_-$ & \multicolumn{3}{|c|}{$5.06\cdot 10^{1}$} \\
\hline
\end{tabular}}
\end{minipage}
\caption{Solutions of the Boltzmann equations for three sample points from the 2CDM case where $\psi_-$ and $\psi_+$ are stable are shown in the upper panels.
The left, middle and right plots are for the values of parameter $\gx=0.1,0.5$ and $5$, respectively. The values of other parameters are shown in the legends of plots and other details are same as in Fig.~\ref{fig:strategy_b}. 
The tables contain first two non-vanishing coefficients of thermally-averaged cross-sections~[pb] expanded in powers of $x^{-1}$, given by $\langle \sigma^{ijkl}\vmol\rangle\! =\! a_Nx^{-N} + a_{N+1}x^{-(N+1)} + \cdots$, and the decay width $\langle \Gamma_{X \rightarrow \psi_+\psi_-}\rangle$~[GeV].}
\label{fig:degenerate_b}
\end{figure*}

Figure~\ref{fig:scan2fdm_pp} shows results of a scan over $\sin\!\alpha, \,\gx,\, m_-$ for fixed $m_{h_2}, m_X$ and $\Delta m= 100\gev$ (left panel)
or $50\gev$ (right panel). All the points  satisfy the correct relic density (for the total abundance) observed by PLANCK at $5\sigma$ and the recent direct detection experimental bound from LUX2016 at $2\sigma$. In this case the $\psi_-$ turns out to be the dominant DM component in most of the parameter space.
The second Higgs boson mass was chosen to be $m_{h_2}=120, 125, 130\gev$ and  $390, 400, 410\gev$ in the left and right panels, respectively.  Therefore the left panel allows for partial cancellation between an exchange of $h_1$ and $h_2$ both for $\psi_\pm$ annihilation diagrams and also for $\psi_\pm$-nuclei scattering process to avoid the direct detection limits even if couplings are not small. Since $50\gev \leq m_- \leq 200\gev$ therefore one can observe both a resonance behavior at $m_-\!\sim \!m_{h_1}/2 \!\sim \! m_{h_2}/2$ in $\psi_\pm\psi_\pm$ annihilation trough s-channel $h_{1,2}$ exchange and also a threshold effect at $m_-\sim m_{h_1} \sim m_{h_2}$ for annihilation into $h_ih_j$ final state. 
The vertical structure observed in Fig.~\ref{fig:scan2fdm_pp} around $m_-\sim 60\gev$ corresponds to a domination of $\psi_\pm\psi_\pm \to h_i^\ast \to VV,\bar{f} f$. As it is seen from the plot large values of $\gx$ are needed, this is a consequence of partial cancellation between $h_1$ and $h_2$ exchange. On the other hand, the independence on $\gx$ could be understood as a result of resonance enhancement around $m_-\! \sim\! m_{h_1}/2\! \sim\! m_{h_2}/2$: even a tiny change of $m_-$ can compensate large variation of $\gx$. The other triple-branch structure that starts around $m_- \sim 120\gev$ corresponds to a threshold for the process $\psi_\pm \psi_\pm \to h_i h_j$. Its initial steepness represents the opening of the $h_i h_j$ final state that must be compensated by suppression of $\gx$ in order to generate correct dark matter abundance.

The right panel of Fig.~\ref{fig:scan2fdm_pp} with its three distinct branches corresponds to a vicinity of the resonance at $m_-\!\sim\! m_{h_2}/2 \!=\! (390, 400, 410)/2\gev$. In this case $\gx$ must be small to compensate the resonance enhancement, therefore direct detection limits are easily satisfied.

In Fig.~\ref{fig:degenerate_b}, we illustrate solutions of the Boltzmann equations for three sample points in the parameter space. The middle panel corresponds to the correct abundance and is in agreement with the LUX upper limits on the DM-nucleon cross-section, which is also present in the scan results shown in the right panel of Fig.~\ref{fig:scan2fdm_pp} as a black star $\star$. As in the previous case of stable $X_\mu$ and $\psi_-$, here we also observe relevance of semi-annihilation and conversion processes and strong $\gx$ dependence. One can see a discrepancy with {\tt micrOMEGAs} results, which are substantial especially in the left panel. The reason for that is the influence of unstable component $X_\mu$, which can be properly described only with a set of 3 coupled Boltzmann equations, whereas in {\tt micrOMEGAs} one has to assume it is in chemical equilibrium with one of the stable components.

\subsubsection{3CDM: a vector and two Majorana fermions as dark matter}
\label{A vector and two Majorana fermions as dark matter}
\begin{figure*}[t]
\centering
\includegraphics[width=0.5\textwidth]{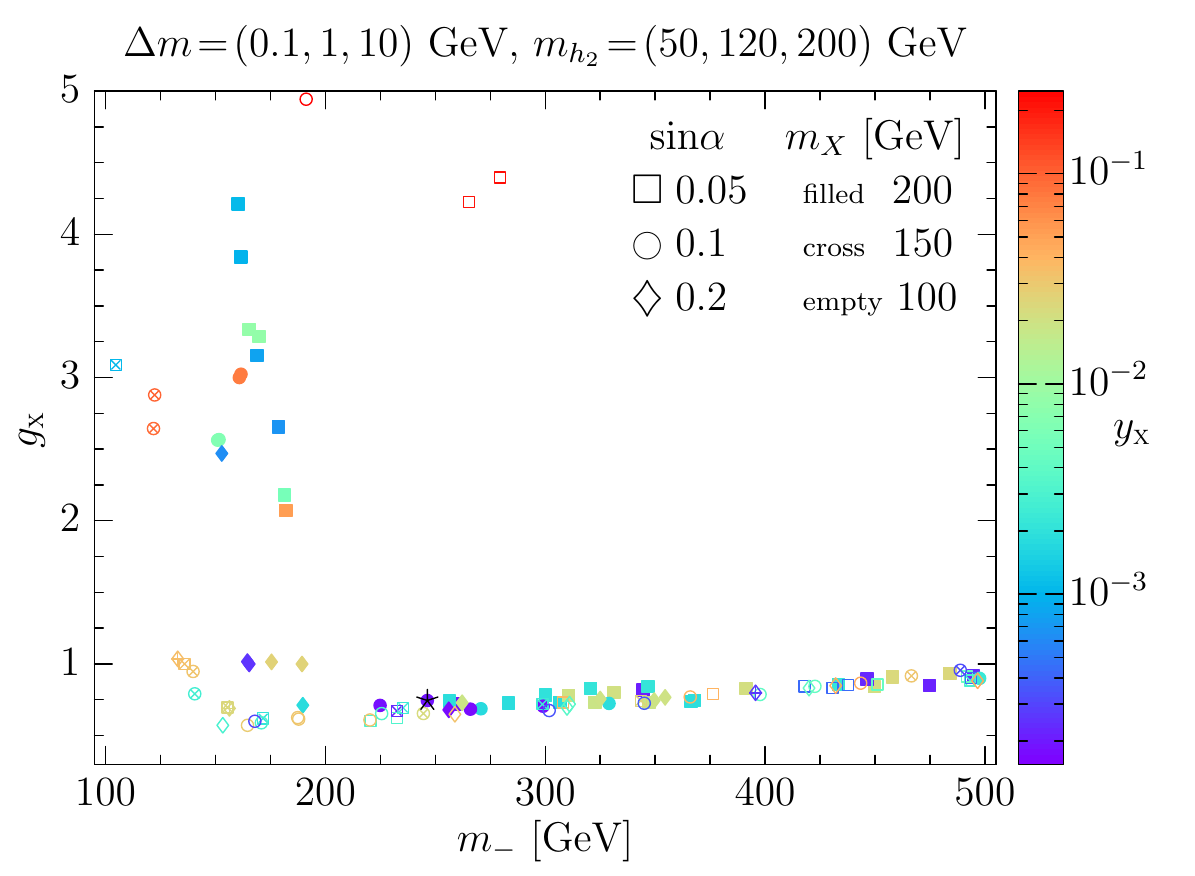}\hspace{-5pt}
\includegraphics[width=0.5\textwidth]{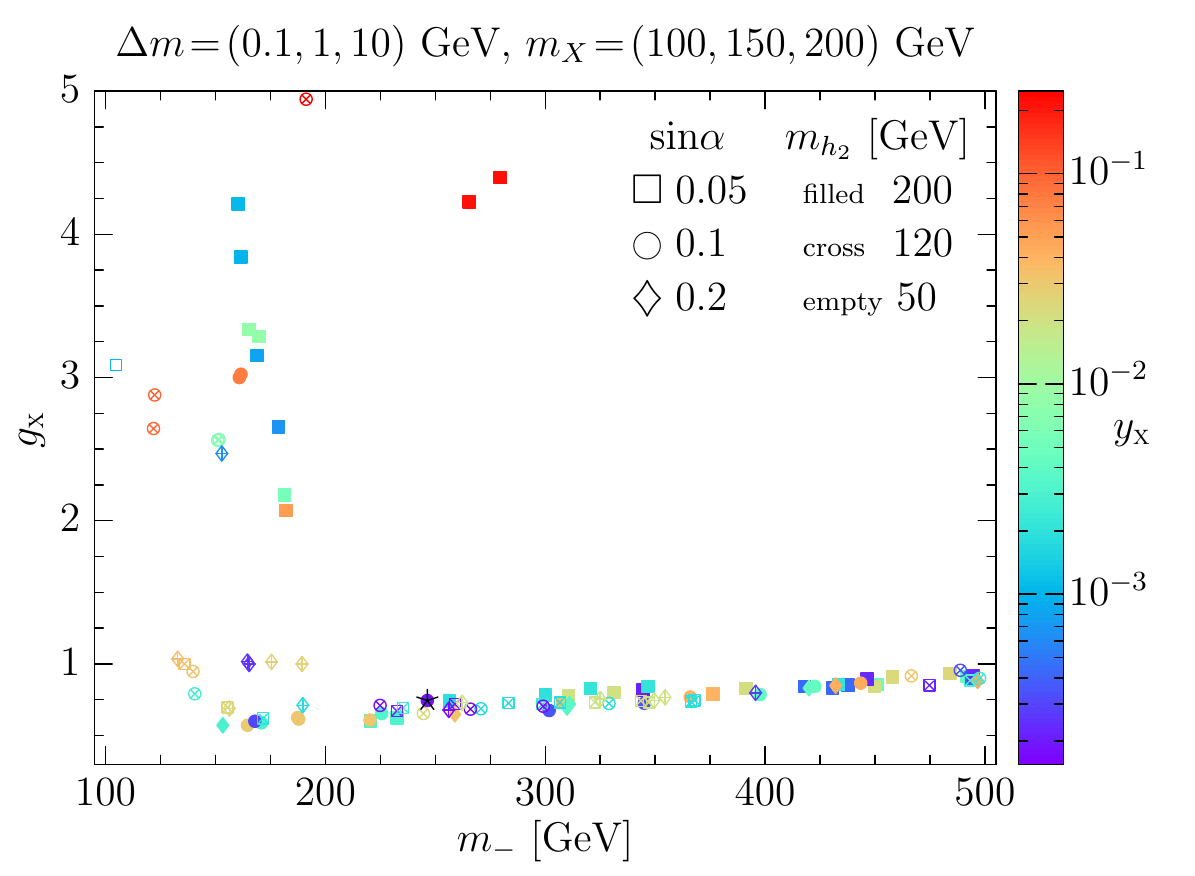}
\caption{These plots show results of scan over the parameter space of a 3CDM case, where $X_\mu$ and $\psi_\pm$ are stable. In this scan we fix three different values of $m_X$, $m_{h_2}$ and $\Delta m$ and vary $m_-$ and $g_x$ as shown in the plots. The left and right plots represent the same data set however points markers are changed from $m_X$ (left) to $m_{h2}$ (right), whereas the color represent the values of dark Yukawa coupling $y_{\textsc x}$.  All the points shown satisfy the correct relic density observed by PLANCK at $5\sigma$ and the recent direct detection experimental bound from LUX2016 at $2\sigma$.
The point denoted by the black star $\star$ in the these plots at $m_-=246.4$~GeV corresponds to the same parameters as those adopted in the middle panel of Fig.~\ref{fig:degenerate_a}.
}
\label{fig:scanvfdm_3com}
\end{figure*}

In this subsection we show results for the scenario with $m_+ +m_->m_X>m_+-m_-$, so all the three dark components are stable. Figure~\ref{fig:scanvfdm_3com} contains results of a scan performed adopting our dedicated code~\!\footnote{Unfortunately {\tt micrOMEGAs} is limited to at most two dark matter components.} that solves the set of the three Boltzmann equations (\ref{boltzmann_X} - \ref{boltzmann_psi_p}).
All the points presented in Fig.~\ref{fig:scanvfdm_3com} satisfy the correct relic density (for the total abundance) observed by PLANCK at $5\sigma$ and the direct detection experimental bound from LUX2016 at $2\sigma$. The scan is performed over $m_-, \; g_x$ with fixed values of $\sin\!\alpha\!=\!(0.05,0.1,0.2)$, $m_X\!=\!(200,150,100) \gev$ and $m_{h_2}\!=\!(200,120,50) \gev$. Note that the left and right panels of Fig.~\ref{fig:scanvfdm_3com} are for the same data-set but for different filling style, in the left and right panel the filling corresponds to $m_X$ and $m_{h_2}$, respectively. 
Here we have tested sensitivity to $\Delta m\equiv m_+-m_-$ focusing on small $\Delta m\!=\!(0.1, 1, 10)$~GeV. 
As it is seen from the Fig.~\ref{fig:scanvfdm_3com}, for $m_-\sim m_+\gsim 200$~GeV relatively small $U(1)_X$ coupling is required, $\gx=0.5-1$, in order to suppress too fast $\psi_\pm\psi_\pm$ s-channel annihilation. Note that
$\yx\!=\! \Delta m\, \gx/(2m_X)$ therefore this annihilation is already quite strongly suppressed by the small Yukawa coupling $\yx$. An important  final state is $t\bar t$, so if $m_-\sim m_+ \lsim m_t$ this annihilation channel closes so that even large $\gx$ is allowed/necessary, as observed in the figure.

\begin{figure*}[t]
\centering
\begin{minipage}[h]{0.33\textwidth}
\includegraphics[width=\textwidth]{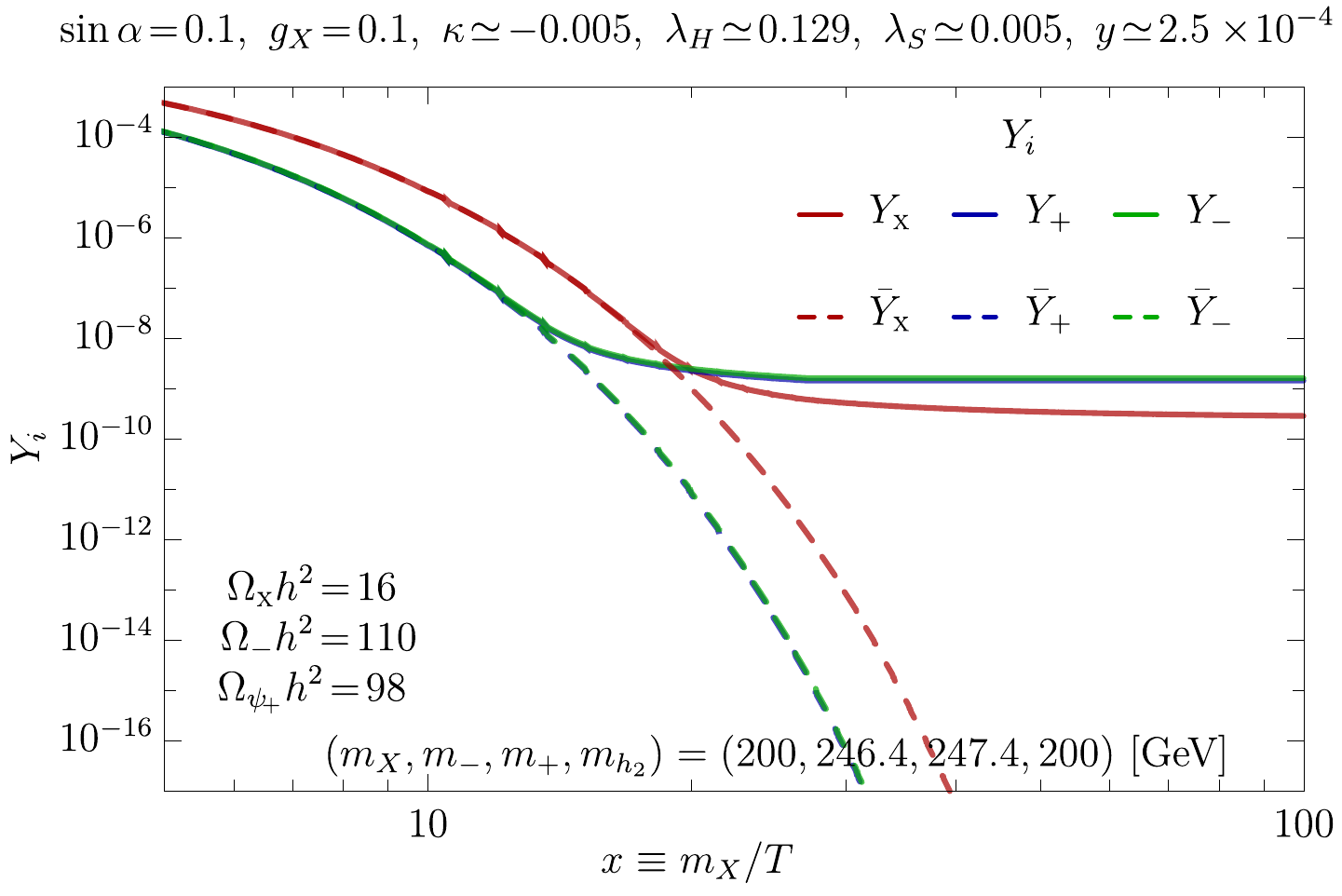}
\resizebox{\textwidth}{!}{%
\begin{tabular}{|c|c|c|c|}
\hline
process & $a_N$ & $a_{N+1}$ & $N$ \\
\hline
$XX \rightarrow \text{SM}$ & $1.3\cdot 10^{-3}$ & $1.43\cdot 10^{0}$ & $0$ \\
$\psi_+\psi_+ \rightarrow \text{SM}$ & $1.8\cdot 10^{-8}$ & $-8.6\cdot 10^{-8}$ & $1$ \\
$\psi_-\psi_- \rightarrow \text{SM}$ & $1.8\cdot 10^{-8}$ & $-9\cdot 10^{-8}$ & $1$ \\
\hline
$\Psi_+\Psi_+ \rightarrow XX$ & $3.6\cdot 10^{-4}$ & $3.2\cdot 10^{-3}$ & $0$ \\
$\Psi_-\Psi_- \rightarrow XX$ & $3.5\cdot 10^{-4}$ & $3.2\cdot 10^{-3}$ & $0$ \\
$\Psi_+,\Psi_+ \rightarrow \Psi_-\Psi_-$ & $6.6\cdot 10^{-4}$ & $4.5\cdot 10^{-2}$ & $0$ \\
\hline
$\Psi_+\Psi_- \rightarrow Xh_1$ & $2.1\cdot 10^{-5}$ & $-7.6\cdot 10^{-5}$ & $0$ \\
$\Psi_+\Psi_- \rightarrow Xh_2$ & $1.5\cdot 10^{-3}$ & $-4.3\cdot 10^{-3}$ & $0$ \\
$X\Psi_- \rightarrow \Psi_+h_1$ & $7.7\cdot 10^{-6}$ & $2.1\cdot 10^{-4}$ & $0$ \\
$\Psi_+h_2 \rightarrow X\Psi_-$ & $4.2\cdot 10^{-5}$ & $2\cdot 10^{-2}$ & $0$ \\
$X\Psi_+ \rightarrow \Psi_-h_1$ & $8.1\cdot 10^{-6}$ & $2.1\cdot 10^{-4}$ & $0$ \\
$X\Psi_+ \rightarrow \Psi_-h_2$ & $3.4\cdot 10^{-6}$ & $5.4\cdot 10^{-3}$ & $0$ \\
\hline
\end{tabular}}
\end{minipage}
\!\begin{minipage}[h]{0.33\textwidth}
\includegraphics[width=\textwidth]{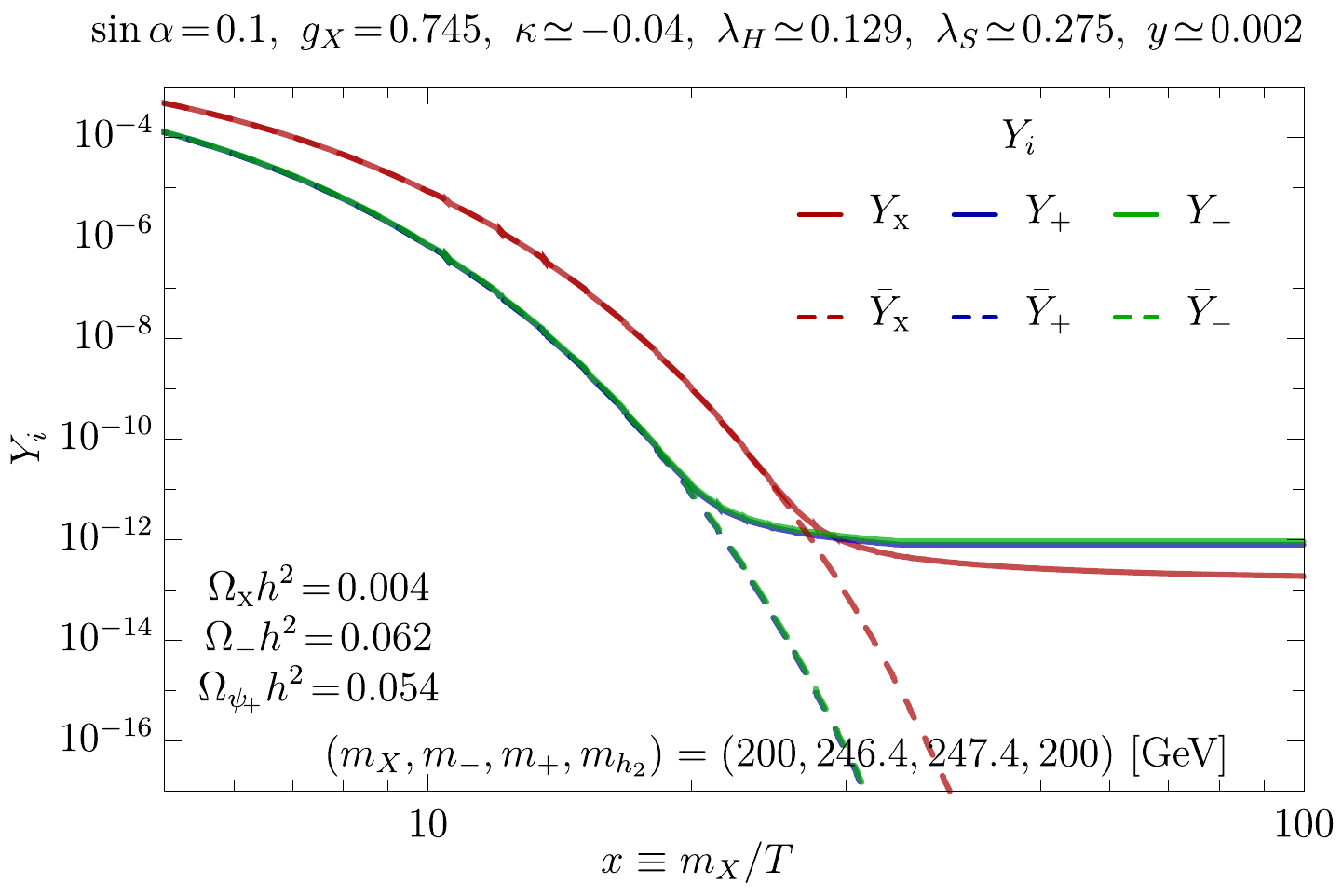}
\resizebox{\textwidth}{!}{%
\begin{tabular}{|c|c|c|c|}
\hline
process & $a_N$ & $a_{N+1}$ & $N$ \\
\hline
$XX \rightarrow \text{SM}$ & $1.55\cdot 10^{0}$ & $4.4\cdot 10^{3}$ & $0$ \\
$\psi_+\psi_+ \rightarrow \text{SM}$ & $4.4\cdot 10^{-5}$ & $-2\cdot 10^{-4}$ & $1$ \\
$\psi_-\psi_- \rightarrow \text{SM}$ & $4.7\cdot 10^{-5}$ & $-2.1\cdot 10^{-4}$ & $1$ \\
\hline
$\Psi_+\Psi_+ \rightarrow XX$ & $1.11\cdot 10^{0}$ & $9.87\cdot 10^{0}$ & $0$ \\
$\Psi_-\Psi_- \rightarrow XX$ & $1.08\cdot 10^{0}$ & $9.95\cdot 10^{0}$ & $0$ \\
$\Psi_+,\Psi_+ \rightarrow \Psi_-\Psi_-$ & $2.03\cdot 10^{0}$ & $1.38\cdot 10^{2}$ & $0$ \\
\hline
$\Psi_+\Psi_- \rightarrow Xh_1$ & $6.6\cdot 10^{-2}$ & $-2.3\cdot 10^{-1}$ & $0$ \\
$\Psi_+\Psi_- \rightarrow Xh_2$ & $4.7\cdot 10^{0}$ & $-1.32\cdot 10^{1}$ & $0$ \\
$X\Psi_- \rightarrow \Psi_+h_1$ & $2.4\cdot 10^{-2}$ & $6.5\cdot 10^{-1}$ & $0$ \\
$\Psi_+h_2 \rightarrow X\Psi_-$ & $1.3\cdot 10^{-1}$ & $6.3\cdot 10^{1}$ & $0$ \\
$X\Psi_+ \rightarrow \Psi_-h_1$ & $2.5\cdot 10^{-2}$ & $6.5\cdot 10^{-1}$ & $0$ \\
$X\Psi_+ \rightarrow \Psi_-h_2$ & $1\cdot 10^{-2}$ & $1.67\cdot 10^{1}$ & $0$ \\
\hline
\end{tabular}}
\end{minipage}
\!\begin{minipage}[h]{0.33\textwidth}
\includegraphics[width=\textwidth]{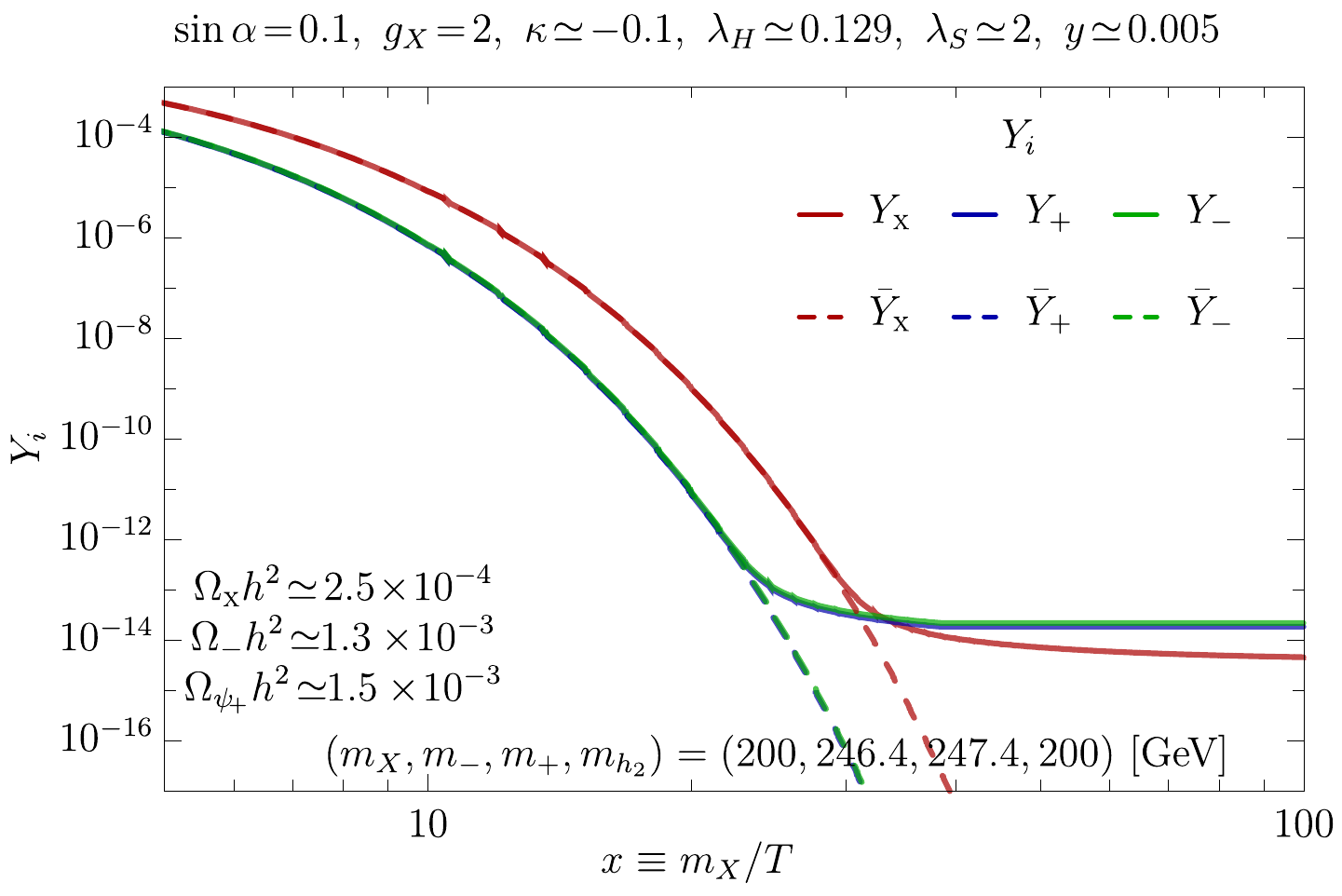}
\resizebox{\textwidth}{!}{%
\begin{tabular}{|c|c|c|c|}
\hline
process & $a_N$ & $a_{N+1}$ & $N$ \\
\hline
$XX \rightarrow \text{SM}$ & $7.89\cdot 10^{1}$ & $2.29\cdot 10^{5}$ & $0$ \\
$\psi_+\psi_+ \rightarrow \text{SM}$ & $2.3\cdot 10^{-3}$ & $-1\cdot 10^{-2}$ & $1$ \\
$\psi_-\psi_- \rightarrow \text{SM}$ & $2.4\cdot 10^{-3}$ & $-1.1\cdot 10^{-2}$ & $1$ \\
\hline
$\Psi_+\Psi_+ \rightarrow XX$ & $5.77\cdot 10^{1}$ & $5.12\cdot 10^{2}$ & $0$ \\
$\Psi_-\Psi_- \rightarrow XX$ & $5.59\cdot 10^{1}$ & $5.16\cdot 10^{2}$ & $0$ \\
$\Psi_+,\Psi_+ \rightarrow \Psi_-\Psi_-$ & $1.05\cdot 10^{2}$ & $7.19\cdot 10^{3}$ & $0$ \\
\hline
$\Psi_+\Psi_- \rightarrow Xh_1$ & $3.41\cdot 10^{0}$ & $-1.22\cdot 10^{1}$ & $0$ \\
$\Psi_+\Psi_- \rightarrow Xh_2$ & $2.44\cdot 10^{2}$ & $-6.93\cdot 10^{2}$ & $0$ \\
$X\Psi_- \rightarrow \Psi_+h_1$ & $1.24\cdot 10^{0}$ & $3.37\cdot 10^{1}$ & $0$ \\
$\Psi_+h_2 \rightarrow X\Psi_-$ & $6.66\cdot 10^{0}$ & $3.27\cdot 10^{3}$ & $0$ \\
$X\Psi_+ \rightarrow \Psi_-h_1$ & $1.29\cdot 10^{0}$ & $3.36\cdot 10^{1}$ & $0$ \\
$X\Psi_+ \rightarrow \Psi_-h_2$ & $5.4\cdot 10^{-1}$ & $8.65\cdot 10^{2}$ & $0$ \\
\hline
\end{tabular}}
\end{minipage}
\caption{Solutions of the Boltzmann equations for three sample points from the 3CDM case where all three dark states ($X,\psi_\pm$) are stable are shown in the upper panels. The left, middle and right plots are for the values of parameter $\gx=0.1,0.745$ and $2$, respectively. The values of other parameters (same in all the panels) are shown in the legends of plots. The tables contain first two non-vanishing coefficients of thermally-averaged cross-sections [pb] expanded in powers of $x^{-1}$, given by $\langle \sigma^{ijkl}\vmol\rangle\! =\! a_Nx^{-N} + a_{N+1}x^{-(N+1)} + \cdots$. Note that for these points the dark fermion Yukawa coupling is very small, i.e. $y\ll1$, as a result the direct annihilation processes for $\psi_\pm$ are inefficient, therefore the main annihilation processes are through the semi-annihilations and conversions to $X$ which further annihilate to SM.}
\label{fig:degenerate_a}
\end{figure*}

Figure~\ref{fig:degenerate_a} shows solutions of the Boltzmann equations for three sample points in the parameter space of 3CDM. The middle panel represents the point marked as a black star~$\star$ in the scan results of Fig.~\ref{fig:scanvfdm_3com} that satisfies relic abundance and LUX2016 constraints. As in the previous cases for 2CDM, here we also observe relevance of semi-annihilation and conversion processes and strong $\gx$ dependence on the yields of dark matter components.
%
\subsection{Limiting cases}
\label{Limiting cases}
In this subsection we are going to discuss special regions in the parameter space of the model that result in simpler, models of DM.
\subsubsection{The vector dark matter (VDM) model}
\label{VDM}
If mass splitting between $\psi_+$ and $\psi_-$ is small comparing to $m_X$, then, for fixed $g_X$, the Yukawa coupling
$y_X$ is suppressed as $y_X=\Delta m/(2 v_x)=(\Delta m/m_X)(g_X/2)$. Therefore, in this limit, since Yukawa couplings become irrelevant,  the model might be reduced to the 1-component VDM model (see e.g. \cite{Duch:2015jta} and \cite{Duch:2017khv} where the same notation as here has been adopted). Note however that even though fermionic DM decouple from the SM, nevertheless it is still present and may influence cosmological dynamics and contribute to the observed amount of DM. In order to enable  efficient  $\psi_\pm$ annihilation it is sufficient to assume that $2 m_D > m_X+m_i$ and/or $m_X > m_i$ ($m_D\equiv (m_++m_-)/2$) so that at present only $X_\mu$ contributes to the observed DM abundance and could be successfully fitted by tuning $g_X$. 
Samples of parameter sets that imply proper $\Omega_{DM}$ and fit in the VDM limit are shown in Tab.~\ref{tab:limitingCases}. 
In ref.~\cite{Duch:2017khv} the VDM has been analyzed focusing on the possibility of enhancing self-interaction, and some regions of the parameters space where elastic $XX$ scattering is amplified and all other constraints are satisfied have been found.
\subsubsection{The fermion dark matter (FDM) model}
\label{FDM}
Another interesting limit of our model is a renormalizable model of fermionic DM, see e.g. \cite{Kim:2008pp,Ettefaghi:2013xi,Baek:2011aa,Baek:2017vzd}. Those models usually employ an extra singlet real scalar field that couples to the SM Higgs doublet via the Higgs portal and to a singlet dark Dirac fermion as well. In the fermionic DM limit of our model, we recover a model of a Majorana singlet DM  that couples to a complex scalar $S$. Since our model is invariant under local $U(1)_X$ therefore in the limit of small gauge coupling, $g_X \ll 1$, and substantial mass-splitting between $\psi_+$ and $\psi_-$ (so for enhanced Yukawa coupling, $y_X$), effectively we obtain a renormalizable model symmetric under a global $U(1)_X$ of a single Majorana dark fermion $\psi_-$ interacting with $S$.  The scalar $S$ controls communication between dark sector and the SM. Examples of parameter sets that imply proper $\Omega_{DM}$ and fit in the FDM limit are shown in Tab.~\ref{tab:limitingCases}. The model is slightly more restrictive than those considered earlier in \cite{Kim:2008pp,Ettefaghi:2013xi,Baek:2011aa,Baek:2017vzd} since our DM is a Majorana fermion and the scalar potential is more restricted, as being invariant under the global $U(1)_X$, however predictions of those models are similar, see \cite{Baek:2011aa}. In order to obtain fermionic (Majorana) DM model one should reduce $g_X$ in order to decouple $X_\mu$, keeping in mind that certain minimal interaction strength is necessary for $X_\mu$ to maintain kinetic equilibrium. Note that in order to reduce the model to a single fermionic DM model, we have to remove somehow $\psi_+$ and $X_\mu$. The easiest way to get rid of $\psi_+$ is to assume that it is the heaviest dark state, so that it will have a chance to decay quickly. If the following mass ordering, $m_+ > m_X > m_-$ is fulfilled, then indeed the dominant DM component is the Majorana fermion $\psi_-$, while other dark components disappear. The model contains, of course, two scalar Higgs bosons that mix in the standard manner and play a role of mediators between the dark sector and the SM.   
Self-interaction in the FDM model has been discussed in \cite{Kainulainen:2015sva}. It turns out that for the self-interactions
to be sufficiently strong, the scalar mediator, $h_2$, has to be very light what implies well-known problems~\cite{Bringmann:2016din} in the early Universe if $h_2$ is present during the era of BBN.
\begin{table}[t]
\caption{Sets of parameters implying limiting VDM and FDM cases with proper $\Omega_{DM}$. Relic densities of $X_\mu$ and $\psi_-$ are provided, density of $\psi_+$ is negligible. All the masses are in GeV.}
\label{tab:limitingCases}
\centering{\tabulinesep=1pt
\begin{tabu}to 0.9\textwidth{|[1pt]X[1,c]|[0.75pt]X[1,c]|[0.75pt]X[1,c]|[0.75pt]X[1,c]|[0.75pt]X[1,c]|[0.75pt]X[1,c]|[0.75pt]X[1.5,c]|[0.75pt]X[1.5,c]|[1pt]}\hline\rowcolor{gray!37}
$m_X$&$m_+$&$m_-$&$m_{h_2}$&$g_X$&$\sin\alpha$&$\Omega_{\textsc{x}}h^2$&$\Omega_-h^2$\\
\hline 
\multicolumn{8}{|c|}{VDM}\\
\hline \rowcolor{gray!13}
$~100~$&$~405~$&$~400~$&$~180~$&$~0.4~$&$0.1$&$0.121$&$1.72\!\cdot\! 10^{-15}$\\
\rowcolor{gray!3}
$200$&$705$&$700$&$120$&$0.256$&$0.1$&$0.121$&$1.61\!\cdot\! 10^{-19}$\\
\hline
\multicolumn{8}{|c|}{FDM}\\
\hline\rowcolor{gray!13}
$100$&$2500$&$19$&$50$&$0.3$&$0.3$&$5.71\!\cdot\! 10^{-4}$&$0.121$\\
\rowcolor{gray!3}
$100$&$5\!\cdot\! 10^{4}$\!&$40$&$140$&$0.1$&$0.25$&$1.20\!\cdot\! 10^{-4}$&$0.120$\\
\hline
\end{tabu}
}
\end{table}
\subsubsection{The fermion dark matter (FDM) model with a stable vector mediator}
\label{stable_vector_med}
Another interesting limit of our model has been considered very recently in \cite{Duerr:2018mbd}. If, in our model, 
$\Delta m \to 0$ then  the Yukawa coupling $y_X$
vanishes and masses of Majorana fermions $\psi_+$ and $\psi_-$ become degenerate. Then our model reduces to the model considered in \cite{Duerr:2018mbd}, which is just a model of a Dirac fermion as a DM and a stable vector mediator. Their~\cite{Duerr:2018mbd} mediator corresponds to our vector component of DM, $X_\mu$, while the dark Dirac fermion is an analog of our degenerate Majorana dark fermions $\psi_+$ and $\psi_-$. The authors of \cite{Duerr:2018mbd} show that the model can indeed predict enhanced DM self-interaction while satisfying all existing experimental constraints if mass of the stable vector mediator is of the order of $1\mev$. This has been also confirmed in the appropriate ($\Delta m \to 0$, $m_X \sim {\cal O}(1) \mev$) region of the parameter space of our model. In this case the DM abundance is dominated by mass-degenerate $\psi_\pm$, even though formally it is a 3-component case (3CDM) if $\Delta m < m_X$.  

\subsection{Distinguishing limiting cases}
\label{distinguishing_cases}

As it has been discussed in the previous subsection the model discussed here simplifies in various regions of the parameter space
where it reduces to a single-component dark matter mode i.e. FDM or VDM. In remaining parts of the parameter space it describes 
a genuine 2 or 3 component dark matter. In this context it is natural to rise the question how could one disentangle those three
possibilities. Some attempts to address this sort of question have already been made in the literature, see e.g. \cite{Ko:2016xwd,Kamon:2017yfx}, where the authors
considered a possibility to disentangle spin, 0, 1/2 or 1 dark matter at $e^+e^-$ future colliders. The VDM model they considered was the same as the  limiting version of our model discussed in sec.~\ref{VDM}, however their FDM was slightly different than ours from sec.~\ref{FDM}. Of course, they did not discuss multi-component scenario.
An exhaustive discussion, that takes into account all existing experimental constraints within a single model that 
allows for 2 or 3 DM components is still missing. Such an analyzes lies beyond the scope of this paper however it shall 
be investigated in the near future~\cite{BGMI}. Nevertheless few comments are here in order.

\bit
\item \underline{Direct detection}\\
Contributions to DM-nucleon scattering consists of the sum of standard $\sigma_{X-N}$ and $\sigma_{\psi_\pm-N}$ cross-sections that 
are not sensitive to the presence of all the 2-3 DM components, rather this is a sum of contributions that exist in 1-component models. 
There exists however a more interesting inelastic scattering process which is sensitive to the multi-component nature of the model considered here, i.e.
$\psi_+ N\to \psi_- X N$, note that all the dark particle are involved, so that this process might provide a signature of the multi-component scenario or perhaps some useful correlation with other observables. This process could be enhanced (and therefore efficiently constrained) for small $m_X$ which on the other hand helps to enhance $\psi_\pm$ self-interaction.    
\item \underline{Indirect detection}\\
Similarly indirect detection experiments,  besides  standard $XX\to {\rm SM}$ and $\psi_\pm \psi_\pm \to {\rm SM}$ contributions receive also more interesting 
one $\psi_+\psi_-\to X h_i$ followed by $h_i$ decays.
\item \underline{Colliders}\\
$e^+e^-$ colliders provide a clean environment that might be used to test the multi-DM scenario considered here. Namely one can investigate the process
$e^+e^-\to Z^* \to h_i Z \to \chi\chi Z$ with $\chi=X$ or $\psi_\pm$. Energy-distribution of $Z$ might be adopted to gain some information on the invisible objects being produced. Initial estimation indicates that in some regions of the parameter space, for sufficiently large luminosity one should be able to disentangle 1- and 2-3 component scenarios.
\eit

\section{Self-interacting DM} 
\label{selfintDM}
It is well known that the cosmological small-scale structure problems, such as the 'cusp vs. core' and the 'too-big-to-fail' problems could be ameliorated if DM self-interaction was sufficiently strong at the dwarf galaxy scale~\cite{deLaix:1995vi,Spergel:1999mh,Vogelsberger:2012ku,Zavala:2012us,Rocha:2012jg, Peter:2012jh,Kaplinghat:2015aga,Tulin:2017ara}, the required value of the cross-section is
\begin{eqnarray}\label{DMSI}
0.1\,\frac{{\rm cm^2}}{\rm g} < \frac{\sigma_T}{m_{\rm DM}} < 10\, \frac{{\rm cm^2}}{\rm g} \,,
\end{eqnarray}
where $\sigma_T\equiv \int d\Omega (1-\cos \theta) d\sigma/d\Omega$ is the so-called momentum transfer cross section between DM particles. However,  DM self-scattering cross-section as large as $\sigma_T/m_{\rm DM} \simeq 10~{\rm cm^2/g}$ turns out to be disallowed by observations at the cluster scale with the typical constraint $\sigma_T/m_{\rm DM} < 1~{\rm cm^2/g}$~\cite{Clowe:2003tk,Markevitch:2003at,Randall:2007ph,Kahlhoefer:2013dca,Harvey:2015hha}. Therefore in the following we will try to find a region in the parameter space where $0.1~{\rm cm^2}/{\rm g}<\sigma_T/m_{\rm DM}<1~{\rm cm^2}/{\rm g}$.
ss
A possible strategy that may generate large DM self-interaction is to introduce a mediator which is much lighter than the DM particles. In the VFDM model, there are two options, the mediator could be either $h_2$ or DM component $X_\mu$. As shown in \cite{Duch:2017khv} the choice of light $h_2$ implies number of severe constraints therefore here we will focus on the case of light vector DM component, which may serve as a mediator in elastic $\psi_+\psi_-$ scattering. The main contribution to the amplitude for  $\psi_+\psi_- \to \psi_+\psi_-$ comes from the t-channel $X_\mu$-exchange.
The transfer cross-section for the two Majorana eigenstates interacting with a vector mediator was discussed in \cite{Zhang:2016dck}. In case of the small mass splitting $m_+-m_- \ll m_D$, we can use the result obtained in a Born approximation for the Dirac fermion \cite{Tulin:2013teo}
\beq
\sigma_T \!\equiv\! \int\!\! d\Omega(1-\cos\theta) \frac{d\sigma}{d\Omega}\!=\! 
\frac{\gx^4}{16\pi m_-^2 v^4}\!\left(\frac{m_+}{m_-}-\frac{1}{2}\right)\!\left[\log\left(1+\frac{m^2_- v^2}{m_X^2}\right)-\frac{m^2_- v^2}{m_X^2+m^2_- v^2}\right]
\label{sig_T}
\eeq
The above perturbative result is valid if $g_X^2 m_-/m_X \lsim 16\pi$ and  $(m_+-m_-)\ll m_D$. The substantial enhancement could be achieved for light $X_\mu$ with its mass e.g. $m_X \sim {\cal O}(1 \mev)$. In addition in case of small mass splitting the contribution from t-channel $h_2$ exchange is suppressed. For $m_- \gg m_X$ also the s-channel $X_\mu$ exchange could be neglected. Here we consider the scenario with three stable components, therefore $m_X > m_+-m_-$.

In the region of parameters considered here abundance of $\psi_\pm$ and $X_\mu$ might be comparable, however for instance domination of $\psi_\pm$ could also be reached by facilitating $XX$ annihilation by assuming $m_2 < m_X$, then appropriate $g_X$ might  be adjusted to tune the proper total DM abundance. Note that the Yukawa coupling $y_X$ remains small so that the potential relevance of $h_2$ mediation is therefore limited.  If $(m_+-m_-)\ll m_D$ then both indirect detection of $\psi_\pm\psi_\pm$ annihilating at present time and the cross-section for $\psi_\pm$-nucleon scattering would be sufficiently suppressed. The mixing angle is as usually assumed to be small, $\sin\theta \ll 1$, what provides additional suppression of both direct and indirect detection. Concluding, it seems that there exists the region of parameter space consistent with the data and providing substantial self-interaction of DM components and large ratio of masses ($m_D \gg m_X$) for DM components. This illustration of possibility for enhanced self-interaction in our model is located in a region of parameter space, which is similar to the limit considered in sec.~\ref{stable_vector_med}, i.e. for fermionic Dirac DM and stable vector mediator discussed very recently in \cite{Duerr:2018mbd}.

\begin{figure}[t]
\centering
\includegraphics[width=0.57\columnwidth]{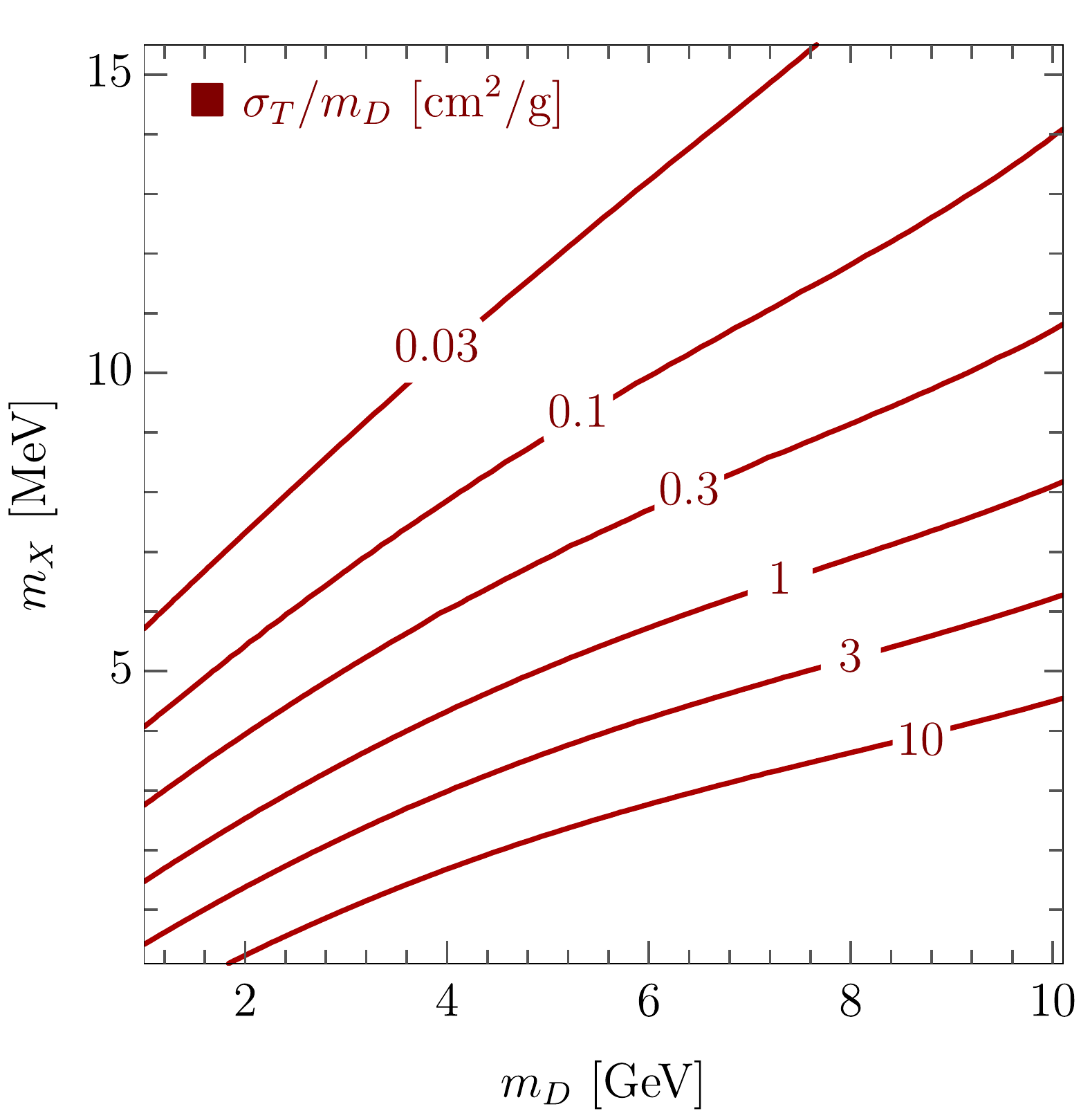}
\caption{Contours of self-interaction cross-section $\sigma_T/m_{D}$ at dwarf galaxies scale ($v=10$~km/s) in the ($m_{D}, m_X$) plane. For each point the value of $g_X$ was fitted using relic density constraint.}
\label{selfContPlot}
\end{figure}

In the Fig.~\ref{selfContPlot}, we show results of detailed scans focused on that region. The relic abundance was calculated using {\tt micrOMEGAs} code~\cite{Belanger:2014vza} by placing $\psi_+$ and $\psi_-$ in one dark sector and $X$ in another. Here we assume that both fermions are kept in equilibrium with each other by the efficient exchange processes $\psi_\pm X \leftrightarrow \psi_\pm h_2$. The scan was made over masses in the range $m_X\in[1,15]$~MeV, $m_D\in[1,10]$~GeV for fixed values of $m_{h_2}\in\{1,2,5\}$~MeV and $\sin\alpha\in\{10^{-5},10^{-6},10^{-7}\}$. Parameter $g_x$ was fitted imposing the condition that density of fermions satisfies relic abundance constraint whereas contribution of $X$ is negligible. The latter is achieved by the effective annihilation $XX\rightarrow h_2h_2$ if $h_2$ is lighter than $X$. We choose the mass splitting $m_+-m_- = 10^{-5}$~GeV. In this way we can ensure that both states are present with nearly the same relic abundance. Moreover as it leads to the suppression of the Yukawa coupling, therefore we can avoid the indirect detection bounds. Another strong constraint comes from the limit on the invisible Higgs decay $h_1\rightarrow h_2 h_2$. It results in the bound $\sin\alpha\leq 10^{-5}$, which on the other hand suppresses the DM-nucleon scattering cross-section to the range which is in agreement with direct detection experiments.

Similar scenario was discussed in the case of Dirac fermion in \cite{Duerr:2018mbd}. Since here we focus on the small mass splitting therefore our model effectively also contains a Dirac dark fermion and a stable vector as in \cite{Duerr:2018mbd}. Therefore, our results for $\sigma_T/m_D$ shown 
in Fig.~\ref{selfContPlot} indeed agree with those obtained in \cite{Duerr:2018mbd}. Note however, this accordance takes place only in this particular region of the parameter space, while in general the models are quite different, for instance, by the presence of Yukawa interactions that are allowed in our model due to specific assignments of dark charges. More comprehensive analysis of our model will be presented elsewhere.

\section{Summary and conclusions}
\label{conclusions}
Multi-component dark matter scenarios are natural extensions of a simple WIMP dark matter. They predict more than one stable component in a dark sector and therefore they constitute a much richer dynamical structure. In this work we have presented a minimal UV-complete vector-fermion DM model with two or three stable particles. Its dynamical properties were discussed.
Our vector-fermion DM model involves a dark sector with a $U(1)_X$ gauge symmetry. The dark matter contents are the dark gauge boson $X_\mu$, a Dirac fermion $\chi$, and a complex scalar $S$, all are charged under the dark $U(1)_X$ gauge symmetry and are neutral under the SM gauge symmetry. Moreover, all the SM particles are neutral under the dark $U(1)_X$ gauge symmetry. The dark sector communicates with the visible sector (SM) through the Higgs portal $\kappa|H|^2|S|^2$. To generate the dark gauge boson $X_\mu$ mass we have employed the Higgs mechanism in the dark sector. 

After the dark sector spontaneous symmetry breaking and mass diagonalization, our vector-fermion DM scenario comprises of a dark vector $X_\mu$ and two dark Majorana fermions $\psi_\pm$.  Out of eight free parameters of the model, the SM Higgs vev $v\!=\!246\gev $ and the SM-like Higgs mass $m_{h_1}\!=\!125\gev$ are fixed which leaves us with six independent parameters. We have chosen the physical basis where the six independent parameters are four masses $m_X,m_\pm, m_{h_2}$, the mixing angle $\sin\!\alpha$, and the dark gauge coupling $\gx$. To guarantee perturbativity we assumed $\gx\!\leq\!4\pi$. We have employed $\sin\!\alpha\leq 0.33$, which is consistent with the $2\sigma$ constraint from current measurements of the SM-like Higgs boson couplings to the SM gauge bosons at the LHC. Our VFDM model has an exact charge conjugation symmetry and the dark gauge symmetry which result in an accidental discrete $Z_2\!\times\!Z_2^\p$ symmetry.
The charge assignments under this $Z_2\!\times\!Z_2^\p$ symmetry are: $X_\mu (-,-),\psi_{+}(-,+), \psi_{-}(+,-)$ and $h_{1,2}(+,+)$ (also all SM gauge bosons and fermions are even under both discrete symmetries). The dynamics of the dark sector is mainly controlled by the gauge coupling $\gx$ which couples the three dark fields, i.e. $X_\mu\bar\psi_{+}\gamma^\mu\psi_{-}$. 

In this work, we have analyzed the dynamics of the dark sector in the thermal freeze-out paradigm by solving the three coupled Boltzmann equations for the dark sector species ($X_\mu,\psi_+,\psi_-$). The vector-fermion DM dynamics turns out to be different in many ways than the standard single component WIMP dark matter scenarios. 
In our model, depending on the masses of the dark sector particle $m_X, m_{+}, m_{-}$ there are the following three distinct cases where either two or all three dark sector particles are stable. 
\bit\itemsep0em
\item[(i)] \underline{$m_+>m_X+m_-$}: a two-component dark matter case where the stable particles are the vector $X_\mu$ and the Majorana fermion $\psi_-$, see sec.~\ref{A vector and a Majorana fermion as dark matter}. In this case we have performed scans over $m_X, \gx$ for different values of $m_\pm,m_{h_2}$ and $\sin\!\alpha$ to search for regions in the parameter space where the dark matter total relic density and current direct detection constraints are satisfied. Importance of the presence of other dark sector states and their interactions, in particular, the semi-annihilations and conversions has been manifested.  Moreover, we have compared the two-component vector-fermion case with the single-component vector dark matter and highlighted the presence of second component, the latter one is especially useful to compensate the under-abundance of the single-component vector dark matter.
\item[(ii)] \underline{$m_X>m_++m_-$}: a two-component dark matter case where the stable particles are the two Majorana fermions $\psi_+$ and $\psi_-$, see sec.~\ref{Two Majorana fermions as dark matter}. In this case we have performed scans over $m_-,\gx$ for different choices of $m_X,\Delta m,m_{h_2}$ and $\sin\!\alpha$ which satisfy the correct total relic abundance and direct detection bounds. As in the previous case, we have highlighted the importance of the presence of more than one stable states in the dark sector and their interactions. In particular, we have illustrated effects of semi-annihilations in Fig.~\ref{fig:degenerate_b}, which are primarily controlled by the single coupling $\gx$.
\item[(iii)] \underline{$m_++m_->m_X>m_+-m_-$}: a three-component dark matter scenario where all three dark sector particles $(X_\mu,\psi_+,\psi_-)$ are stable, see sec.~\ref{A vector and two Majorana fermions as dark matter}. As in the two-component DM cases we have performed scan over $m_-, \gx$ for different choices of $m_X,\Delta m,m_{h_2}$ and $\sin\!\alpha$. To demonstrate the importance of three stable dark matter particles, we have illustrated in Fig.~\ref{fig:degenerate_a} the case with two Majorana fermions nearly degenerate in mass, i.e. $y\ll 1$, hence their standard annihilations are suppressed, due to small Yukawa couplings, and the semi-annihilations are most important for their annihilations.
\eit
Note that all the points presented in our scans satisfy the total relic density $\Omega_{\rm tot}h^2$ at $5\sigma$ as observed by PLANCK and also the $2\sigma$ direct detection bound from LUX2016. Moreover, to understand the dark matter dynamics we have shown the evolution of the yields of dark matter components for each of the above cases for selected benchmark points, supplemented by tables containing all cross-sections for processes involved in collision terms. Also we compared our results for two-component cases with those obtained from the {\tt micrOMEGAs} code~\cite{Belanger:2014vza} and satisfactory agreement has been found, see Figs.~\ref{fig:strategy_b}~and~\ref{fig:degenerate_b}.

We have also discussed limiting cases of the model that are realized in appropriate regions of the parameter space. One of them corresponds to a model with Dirac fermion DM and stable vector mediator, this is an interesting scenario. A possibility of self-interacting DM has also been addressed and the region of parameter space where $\sigma_T/m_{\rm DM}$ can be substantially enhanced has been found.

To summarize, the absence of any direct, indirect or collider signatures of dark matter suggests a direction that leads beyond the single component WIMP-like dark matter. In particular, multi-component dark matter scenarios offer very rich dynamical structures which could solve current dark matter puzzles. In this work, we have presented a minimal  renormalizable vector-fermion dark matter model where the presence of gauge symmetry and charge conjugation in the hidden sector implies the existence of two or three stable (vector and/or Majorana fermions) dark matter particles. The dynamics of the dark sector in our model is primarily controlled by a single parameter, the dark gauge coupling $\gx$ through the interaction $\big(\bar \psi_{\!+} \gamma^\mu\psi_{\!-} - \bar \psi_{\!-} \gamma^\mu\psi_{\!+}\big) X_\mu$ which connects all dark sector states $X_\mu$, $\psi_+$ and $\psi_-$. Such an interaction allows semi-annihilation and decay processes within the dark sector. We have explored the parameter space of our two/three-component VFDM scenarios requiring the correct total relic density and compliance with the current direct detection bounds. 

\section*{Acknowledgements}
We would like to thank Da Huang for many useful discussions. The research of AA has been supported by the Cluster of Excellence {\it Precision Physics, Fundamental Interactions and Structure of Matter} (PRISMA-EXC1098) and grant 05H12UME of the German Federal Ministry for Education and Research (BMBF). MD acknowledges support of the National Science Center (Poland), research project no. 2017/25/N/ST2/01312.
This work has been supported in part by the National Science Centre (Poland) under research projects 2014/15/B/ST2/00108 and 2017/25/B/ST2/00191. 

\appendix

\section{Boltzmann equations for multi-component dark matter}
\label{Boltzmann equations for semi-annihilating dark matter}
We review here the derivation of the Boltzmann equation for the evolution of number density for the multi-component dark matter in the homogeneous and isotropic universe. Let us consider a generic dark sector which allows interactions with the visible sector and the dark sector, i.e., annihilations (co-annihilations), semi~annihilations, conversions, and decays. We can write the Boltzmann equation for  $\chi_i$th particle as
\beq
\frac{dn_i}{dt}+3H n_i= {\cal C}_i+{\cal D}_i,        \label{boltzmanneq}
\eeq
where $H$ is the Hubble expansion parameter, whereas, ${\cal C}_i$ and ${\cal D}_i$ are collision and decay terms for the $\chi_i$th particle with the visible sector and the dark sector. In general, the collision term ${\cal C}_i$ may involve $2\to 2$, $3\to 2$ and similar scattering processes, however, in order to keep the discussion simple, we focus only on $2\to 2$ processes. We assume that all the dark components have the same temperature as the thermal bath. Similarly, the decay term ${\cal D}_i$ may involve more than $2$-body decays however we limit our-self to 2-body decays, the generalization is straightforward. For $2\to 2 $ scattering processes and the 2-body decay processes, we can write down the collision and the decay terms as follows, 
\beq
{\cal C}_i =\sum_{j,k,l} {\cal C}_{ij\to kl},    \lsp     {\cal D}_i =\sum_{j,k} {\cal D}_{i\to jk},    \label{col-dec}    
\eeq
where the summation is over all possible interactions of $\chi_i$ with the visible sector as well as dark-sector particles. 
The collision ${\cal C}_{ij\to kl}$ and the decay ${\cal D}_{i\to jk}$ terms read:
\begin{align}
{\cal C}_{ij\to kl}=-&\!\int\! d\Pi_id\Pi_jd\Pi_kd\Pi_l (2\pi)^4\delta^4(p_i+p_j-p_k-p_l)     \notag\\ 
&\times\!\Big[|{\cal M}_{ij\to kl}|^2f_i f_j(1\pm f_k)(1\pm f_l)-|{\cal M}_{kl\to ij}|^2 f_k f_l(1\pm f_i)(1\pm f_j)\Big], \label{Cijkl}\\
{\cal D}_{i\to jk}=-&\!\int\! d\Pi_id\Pi_jd\Pi_k (2\pi)^4\delta^4(p_i-p_j-p_k)     \notag\\
&\times\!\Big[|{\cal M}_{i\to jk}|^2f_i(1\pm f_j)(1\pm f_k)-|{\cal M}_{jk\to i}|^2f_j f_k(1\pm f_i)\Big], \label{Dijk}
\end{align}
where the phase space integrand is,
\beq 
d\Pi_i=\frac{d^3p_i}{(2\pi)^3 2E_i},
\eeq
with $|{\cal M}_{ij\to jk}|^2$ and $|{\cal M}_{i\to jk}|^2$ being the matrix element squared {\it summed over initial and final spins} for the reaction  $ij \to kl$ and $i\to jk$, respectively. Above the factors of the form $(1\pm f_i)$ are due to the spin statistics, the plus sign for bosons and the minus sign for fermions. Here $f_i$ denotes the distribution function of a given kind of particles, connected with the number density as follows:
\beq\label{eq:n}
n_i=g_i\int\frac{d^3p}{(2\pi)^3}f_i(E,T),
\eeq
with $g_i$ being the number of spin degrees of freedom.\\
Hereafter it is assumed that appropriate symmetry factors for initial~\cite{Gondolo:1990dk} and final~\footnote{With a factor $1/S_f$, where the final state symmetry factor $S_f=\Pi_{n=1}^{n=N}m_n !$ accounts for $N$ groups of identical final state particles of multiplicity $m_n$.} states are included in $|{\cal M}|^2$. 

We adopt the following assumptions:
\bit\itemsep0em
\item 
Time reversal (T) invariance holds, so the amplitudes satisfy, ${\cal M}_{ij\to kl}={\cal M}_{kl\to ij}$ and ${\cal M}_{i\to jk}={\cal M}_{jk\to i}$ ,
\item 
$m \gg T$, $(m_i-\mu_i)/T \gg 1$ (where $m$ is the mass of dark matter species, $T$ is temperature, and $\mu_i$ is the chemical potential), so that the Bose-Einstein (for bosons) and the Fermi-Dirac (for fermions) distribution functions could be approximated by the Maxwell-Boltzmann distribution functions,
\item 
In the absence of quantum degeneracies (which is assumed since the particles form a very dilute gas), in (\ref{Cijkl}-\ref{Dijk}) the blocking and stimulated emission factors can be neglected, so $1 \pm f_i\simeq 1$ will be adopted,
\item
The initial chemical potentials are negligible,
\item 
Standard Model particles are in thermal equilibrium with the thermal bath,
\item 
Scattering processes with the thermal bath enforce kinetic equilibrium (also after decoupling and out of chemical equilibrium), so that phase-space distribution functions for particles involved in the collision satisfy~\cite{Dodelson:2003ft}
\beq
f_i(E,T)=\frac{n_i(T)}{\bar n_i(T)}\times\bar f_i(E,T),    \label{fi_fieq}
\eeq
where $\bar f_i(E)$ is the thermal Maxwell-Boltzmann equilibrium distribution function for zero chemical potential and
\beq
f_i(E,T)=e^{(-E+\mu_i)/T}=e^{\mu_i/T}\bar f_i(E,T),
\eeq
\beq
n_i(T) = g_i e^{\mu_i/T} \int\frac{d^3p_i}{(2\pi)^3}\bar f_i(E)=e^{\mu_i/T} \bar n_i(T).
\eeq
\eit
With the above assumptions we can rewrite the above collision term as,
\begin{align}
{\cal C}_{ij\to kl}\!&=\!-\!\!\int\! \!d\Pi_id\Pi_jd\Pi_kd\Pi_l (2\pi)^4\delta^4(p_i\!+\!p_j\!-\!p_k\!-\!p_l)	\notag\\
& \quad \times\!\big\vert{\cal M}_{ij\to kl}\big\vert^2\Big[\frac{n_i n_j}{\bar n_i\bar n_j}\bar f_i \bar f_j-\frac{n_k n_l}{\bar n_k\bar n_l} \bar f_k \bar f_l\Big]. \label{Cijkl_a}
\end{align}
The thermal  equilibrium distributions satisfy the following relation due to the conservation of energy, 
\beq
\bar f_i \bar f_j=e^{-(E_i+E_j)/T}=e^{-(E_k+E_l)/T}=\bar f_k\bar f_l.
\eeq
After performing the integration over the outgoing momenta, the collision term can be written as, 
\begin{align}
{\cal C}_{ij\to kl}&=-\langle \sigma^{ijkl}\vmol\rangle\Big[n_i n_j-{n_k n_l}\frac{\bar n_i\bar n_j}{\bar n_k\bar n_l}\Big], \label{Cijkl_sigmav}
\end{align}
where $\vmol$ is the M\o ller velocity
\beq
\vmol=\frac{\sqrt{(p_i\cdot p_j)^2-m_i^2m_j^2}}{E_iE_j},
\label{vmol}
\eeq
and the total cross-section {\it summed over initial and final spins}
\begin{align}
&\sigma^{ijkl}(p_i,p_j)=\frac{1}{4E_iE_j v_{\text{M\o l}} }\int d\Pi_kd\Pi_l	\, (2\pi)^4\delta^4(p_i+p_j-p_k-p_l)\, |{\cal M}_{ij\to kl}|^2.
\end{align}
The thermally averaged cross section is defined as, 
\beq
\langle \sigma^{ijkl}\vmol \rangle\equiv \frac{1}{\bar n_i\bar n_j}\int  \frac{d^3p_i}{(2\pi)^3} \frac{d^3p_j}{(2\pi)^3} \,
\sigma^{ijkl}(p_i,p_j) \,\vmol \,\bar f_i\bar f_j,
\eeq
where the equilibrium number density $\bar n_i$ is defined as, 
\beq
\bar n_i\equiv g_i\int\frac{d^3p_i}{(2\pi)^3}\bar f_i({\bf p}).    \label{nieq}
\eeq
After integrations over the momenta and changing variables we can rewrite the above equations as 
\begin{align}
\langle \sigma^{ijkl}\vmol\rangle(x_i) &= \frac{m_i}{8\pi^4 x_i \bar{n}_i\bar{n}_j}\int_{(m_i+m_j)^2}^\infty \!\!\! ds\, \sqrt{s} \, K_1\!\Big(\frac{x_i\sqrt{s}}{m_i}\Big)  p_{ij}^2(s)\, g_{i} g_{j} \,
 {\bar\sigma}_{ij\to kl}(s),
\label{sigmaV}\\
\bar{n}_i(x_i)&=\frac{g_{i}}{2\pi^2 } \frac{m^3_{i} }{x_i}  K_2\left(x_i\right), \hsp\text{and} \hsp x_i\equiv m_i/T,
\end{align}
where ${\bar\sigma}_{ij\to kl}(s)$ is the total cross-section {\it averaged over initial and summed over final spins} ($=\sigma^{ijkl}/(g_ig_j)$) while
$K_{1,2}$ are the Bessel functions of second kind and $p_{ij}^2$ is a square of the incoming particle momenta in the center of mass frame, 
\beq
p_{ij}^2(s)=\frac{\big[s-(m_i+m_j)^2\big]\big[s-(m_i-m_j)^2\big]}{4s}~,
\eeq
which is related to the M\o ller velocity (\ref{vmol}) by
\beq
p_{ij}=\frac{E_iE_j}{\sqrt{s}}\vmol~.
\eeq

Following similar steps as above for calculating the ${\cal C}_{ijkl}$ function \eqref{Cijkl}, one can calculate the ${\cal D}_{i\to jk}$ function \eqref{Dijk}. Assuming T invariant amplitudes for the decaying process, ${\cal M}_{i\to jk}={\cal M}_{jk\to i}$ and using the fact that thermal  equilibrium distributions satisfy the following relation, 
\beq
\bar f_i =e^{-E_i/T}=e^{-(E_j+E_k)/T}=\bar f_j\bar f_k,
\eeq
we can rewrite the decay contribution as,
\begin{align}
{\cal D}_{i\to jk}&=-\int d\Pi_id\Pi_jd\Pi_k (2\pi)^4\delta^4(p_i-p_j-p_k)\, |{\cal M}_{i\to jk}|^2 \,
\frac{\bar f_i}{\bar n_i} \Big[n_i - \bar n_i\frac{n_j n_k}{\bar n_j\bar n_k} \Big]. \label{Dijk_1}
\end{align}
The decay width is defined as usually
\beq
\Gamma_{i\to jk}=\frac{1}{2 m_i} \int d\Pi_j d\Pi_k (2\pi)^4 \delta^4(p_i-p_j-p_k) \,|{\cal M}_{i\to jk}|^2 \, ,
\label{dec}
\eeq
where $|{\cal M}_{i\to jk}|^2$ is a matrix element squared {\it summed over initial and final spins}.
After performing the integration over the outgoing momenta, 
\begin{align}
{\cal D}_{i\to jk}&=-\langle \Gamma_{i\to jk}\rangle\Big[n_i-\bar n_i\frac{n_j n_k}{\bar n_j\bar n_k}\Big]. 
\label{Dijk_sigmav}
\end{align}
where thermally averaged decay rate $\langle \Gamma_{i\to jk}\rangle$ is defined as, 
\beq
\langle \Gamma_{i\to jk}\rangle\equiv \frac{1}{\bar n_i}\int \frac{d^3p_i}{(2\pi)^3} \frac{m_i}{E_i}\Gamma_{i\to jk} \bar f_i = \frac{K_1(x_i)}{K_2(x_i)}\bar\Gamma_{i\to jk},     
\label{decay_ave}
\eeq
where $K_{1,2}$ are the Bessel functions and  $\bar\Gamma_{i\to jk}$
is the width {\it averaged over the initial and summed over final spins}, i.e. $\bar\Gamma_{i\to jk}\equiv\Gamma_{i\to jk}/g_i$.

\section{Direct detection of multi-component dark matter}
\label{DD_multi_DM}
In the MCDM scenario, the standard direct detection bounds given by experimental groups in terms of DM-nucleon scattering cross-section and DM mass cannot be imposed, unless one of the components is responsible for nearly all recoil events \cite{Profumo:2009tb,Dienes:2012cf,Arcadi:2016kmk}. Furthermore, the combined differential event rate in multicomponent case may have a distinctive shape, which allows to discriminate it from single-component scenario \cite{Herrero-Garcia:2017vrl}.  In a general case, one has to confront the theoretical predictions with the results of experiments to put a constraint on the parameter space of the model. In our analysis, we follow \cite{Geng:2016uqt}. The differential recoil event rate for a given DM component $i$ can be written as \cite{Lewin:1995rx}
\beq
\frac{dR_i}{dE_R}=\frac{\sigma_{i N}\rho_i}{2 m_i \mu^2_{iN}} F^2(E_R)\,\eta_i (E_R),
\eeq
where $\rho_i = \Omega_i/\Omega_{\rm tot}\times 0.3 {\rm GeV}/{\rm cm}^3$ is the local density of that DM component, $\sigma_{i N}$ is its nucleus scattering cross-section, $\mu_{iN}$ is the reduced mass of DM-nucleus system, $F$ is the nuclear form factor, which we take as the conventional Helmi form and the function  $\eta_i$ is a mean inverse speed of the DM particles in the local earth frame
\beq
\eta(E_R)=\int_{|v|>v_{\rm min}}\!\! d^3\mathbf{v}\,\frac{f(\mathbf{v})}{v}.
\eeq
For the velocity distribution $f_G(v)$ in our Galaxy we use a truncated Maxwell-Boltzmann  distribution with $v_{\rm esc}=550\;{\rm km}/{\rm s}$.
\beq
f_G (\mathbf{v})= \frac{1}{N_{\rm esc}(\pi v^2_0)^{3/2}}\,e^{-\mathbf{v}^2/v^2_0}\,\theta(v_{\rm esc}-v),
\eeq
where $v_0 = 220 \;{\rm km}/{\rm s}$ is the mean DM velocity relative to galaxy and $N_{\rm esc}$ is the normalization factor. The distribution of DM as observed from the Earth takes into account its velocity $\mathbf{v}_{\rm e}$ relative to the galactic halo rest frame
\beq
f(\mathbf{v}) = f_G(\mathbf{v}+\mathbf{v}_{\rm e}).
\eeq
The total $dR/dE_R$ differential recoil event rate is obtained by summing the rates for all DM components.

Various DM direct detection experiments measure different kinds of detection signals, eg. prompt scintillation signal $S1$, ionization charge signal $S2$, the electron equivalent energy or energy released in photons. To put a constraints on a region of DM parameter space, one has to compute the expected experimental signal from the recoil event rate $dR/dE_R$ of multicomponent DM  obtained above. We focus on the predictions for the $S1$ signal measured by LUX experiment \cite{Akerib:2016vxi}. Following \cite{Geng:2016uqt} we count the number of events $N$ in the signal range $S1\in[S1_a,S2_b]$ as described in \cite{Aprile:2011hx} 
\begin{align}
&N_{[S1_a,S1_b]} =  Ex \int_{S1_a}^{S2_b} dS1 \bigg[\sum_{n=1}^\infty \epsilon(S1){\rm Gauss}(S1|n,\sigma) \int_0^\infty dE_R {\rm Poiss}(n|\nu(E_R))\epsilon_{S2}(E_R)\frac{dR}{dE_R}\bigg),
\end{align}
where additional $S2$ efficiency $\epsilon_{S2}(E_R) \!=\! \theta(E_R - 3{\rm keV})$ is cutting the recoil energies from below and $\nu(E_R)\!=\!g_1 L_y E_R$ is the averaged expected number of photoelectrons from a given recoil event, which is calculated based on LUX gain factor $g_1 = 0.0985$ and photon yield $L_y$ adopted from the middle plot of Fig.~1. in~\cite{Akerib:2015rjg}. The Poisson distribution gives the probability of obtaining $n$ photoelectrons, which in the detector produce signal $S1$ normally distributed around $n$ with $\sigma\! =\! \sqrt{n(\sigma^2_{\rm PMT}+g_1)}$, where $\sigma_{\rm PMT}$ is the single-photon resolution \cite{Akerib:2016mzi}.  We include also the detector efficiency for events passing analysis selection criteria $\epsilon(S1)$,  taken as a black curve from Fig.~2 in  \cite{Akerib:2016vxi}, and calculate the expected signal taking into account the total exposure $Ex = 4.47\times 10^4$ ${\rm kg}\times{\rm days}$.

We assume that all candidate events observed by LUX agree with the background-only model, using $S1_a=1$ and $S1_b=50$ we constraint the number of events in the range $N_{[1,50]}<3.09$ at $95\%$~C.L. (2$\sigma$) based on the Poisson statistics.

\providecommand{\href}[2]{#2}\begingroup\raggedright\endgroup

\end{document}